\documentclass[sigconf]{acmart}

\AtBeginDocument{%
  }

\usepackage{xspace}
\usepackage{bm}
\usepackage{graphicx}
\usepackage{comment}
\usepackage{caption}
\usepackage{subcaption}

\usepackage{enumitem}

\newcommand{\Semipinch}{\textsc{SemiPinch}\xspace}
\newcommand{\Fullpinch}{\textsc{FullPinch}\xspace}
\newcommand{\Doublepinch}{\textsc{DoublePinch}\xspace}
\newcommand{\Voice}{\textsc{Voice}\xspace}

\newcommand{\Dwell}{\textsc{gD}\xspace}
\newcommand{\Pinch}{\textsc{gP}\xspace}
\newcommand{\VSelect}{\textsc{gV}\xspace}

\newcommand{\qmode}{\textsc{Quasi}\xspace}
\newcommand{\pmode}{\textsc{Persistent}\xspace}

\copyrightyear{2026}
\acmConference[CHI '26]{CHI '26: The ACM CHI Conference on Human Factors in Computing Systems}{April 13--April 17, 2025}{Barcelona, Spain}
\acmDOI{10.1145/3772318.3790513}

\acmSubmissionID{6019}

\begin{document}

\title{Eyes on Many: Evaluating Gaze, Hand, and Voice for Multi-Object Selection in Extended Reality}

\author{Mohammad Raihanul Bashar}
\orcid{0000-0002-5271-457X}
\affiliation{
    \institution{Department of Computer Science \& Software Engineering\\Concordia University}
    \city{Montreal}
    \state{Quebec}
    \country{Canada}
}
\email{mohammadraihanul.bashar@mail.concordia.ca}

\author{Aunnoy K Mutasim}
\orcid{0000-0002-5321-7292}
\affiliation{
    \institution{School of Interactive Arts + Technology (SIAT)\\Simon Fraser University}
    \city{Vancouver}
    \state{British Columbia}
    \country{Canada}
}
\email{amutasim@sfu.ca}

\author{Ken Pfeuffer}
\orcid{0000-0002-5870-1120}
\affiliation{
    \institution{ Department of Computer Science\\Aarhus University}
    \city{Aarhus}
    \state{}
    \country{Denmark}
}
\email{ken@cs.au.dk}

\author{Anil Ufuk Batmaz}
\orcid{0000-0001-7948-8093}
\affiliation{
    \institution{Department of Computer Science \& Software Engineering\\Concordia University}
    \city{Montreal}
    \state{Quebec}
    \country{Canada}
}
\email{ufuk.batmaz@concordia.ca}

\renewcommand{\shortauthors}{Bashar et al.}

\begin{abstract}

Interacting with multiple objects simultaneously makes us fast. A pre-step to this interaction is to select the objects, i.e., multi-object selection, which is enabled through two steps: (1) toggling multi-selection mode --- mode-switching --- and then (2) selecting all the intended objects --- subselection. In extended reality (XR), each step can be performed with the eyes, hands, and voice. To examine how design choices affect user performance, we evaluated four mode-switching (\Semipinch, \Fullpinch, \Doublepinch, and \Voice) and three subselection techniques (Gaze+Dwell, Gaze+Pinch, and Gaze+Voice) in a user study. Results revealed that while \Doublepinch paired with Gaze+Pinch yielded the highest overall performance, \Semipinch achieved the lowest performance. Although \Voice-based mode-switching showed benefits, Gaze+Voice subselection was less favored, as the required repetitive vocal commands were perceived as tedious. Overall, these findings provide empirical insights and inform design recommendations for multi-selection techniques in XR.

\end{abstract}


\begin{teaserfigure}
\centering
  \includegraphics[width=0.85\textwidth]{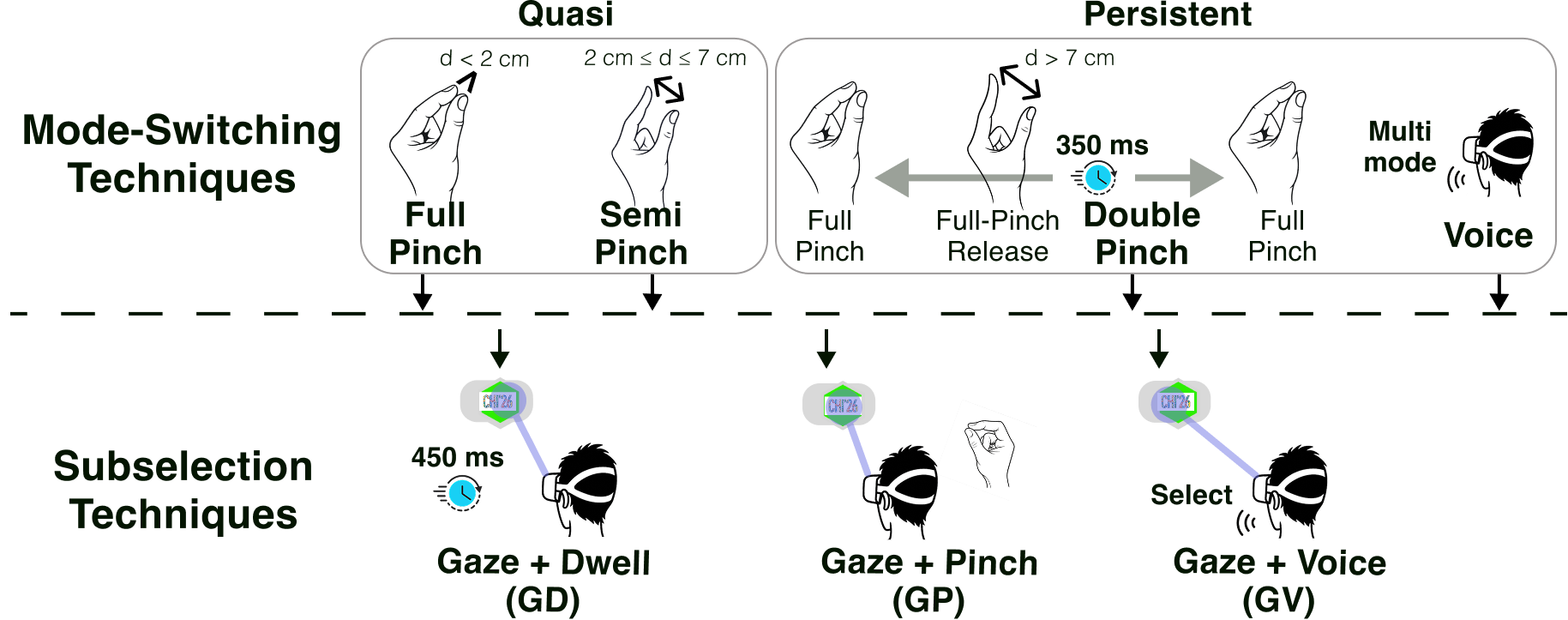}
  \caption{Overview of the mode-switching and subselection techniques for multi-selection in extended reality (XR). We used three pinch states based on the distance (d) between the index and thumb fingertips: Full-Pinch (d~$<$~2~cm), Semi-Pinch (2 cm $\leq$ d $\leq$ 7 cm), and Release (d~$>$~7 cm).  We used two mode-switching strategies --- \qmode (\Fullpinch and \Semipinch) and \pmode (\Doublepinch and \Voice). For subselection, we used Gaze+Dwell (\Dwell), Gaze+Pinch (\Pinch), and Gaze+Voice (\VSelect).
  }
  \Description{Overview of the four mode-switching and three subselection techniques for multi-selection in extended reality (XR). We used two mode-switching strategies --- Quasi (FullPinch and SemiPinch) and Persistent (DoublePinch and Voice). The pinch-based techniques rely on three hand states defined by the distance (d) between the index and thumb fingertips. Full-Pinch is detected when the distance is less than 2~cm, Semi-Pinch when the distance is between 2-7 cm, and Full-Pinch Release when the distance exceeds 7 cm. Double-Pinch is defined as two consecutive full-pinches within 350 ms, while Voice enables mode-switching through predefined voice commands. For subselection, we used Gaze+Dwell (GD; 450 ms), Gaze+Pinch (GP), and Gaze+Voice (GV).}
  \label{fig:teaser}
\end{teaserfigure}


\maketitle

\section{Introduction}

Multi-object selection techniques enable faster manipulation of multiple items together, such as files, images, or documents. While multi-object selection is well understood in 2D interfaces, the spatial interaction afforded by extended reality (XR) introduces new challenges~\cite{lucas2005design, wu2023point}. Different from desktop or touchscreen, many XR systems are designed to work without controllers, and instead, interaction is supported through modalities such as hands, (eye) gaze, and voice. This opens up a large design space. Yet, little is known about how the two fundamental components of multi-selection --- \textit{mode-switching} (i.e., changing between single- or multi-selection states) and \textit{subselection} (i.e., selecting targets to group together) --- and their combinations affect users' performance in XR.

Two types of mode-switching approaches are employed in the literature: \textit{quasi-} and \textit{persistent-mode}. In quasi-mode, users must continuously hold an ``action'' to maintain multi-selection, e.g., holding the SHIFT/CTRL key on a keyboard or using (two-finger) tap-(hold)-and-drag on a touchscreen. Quasi-mode can improve mode awareness (i.e., users better remember which mode they are in), but often increases fatigue, and is thus less suitable for longer tasks~\cite{hinckley2006springboard}. In contrast, persistent-mode is explicitly triggered and released, such as toggling a button (e.g., CAPS LOCK), or tapping-and-holding the first desired object on an Android touchscreen~\cite{dehmeshki2010design, ramanathan2016multi, pfeuffer2014gaze}. Persistent-mode frees up degrees of freedom, allowing users to focus on subselection rather than sustaining a quasi-mode action.

Translating such 2D practices into XR is not straightforward. For instance, quasi-modes with hand gestures can become sustained mid-air hand poses that are fatiguing~\cite{laviola20173d, jang2017modeling} and prone to tracking instability~\cite{bashar2025effect}. 
In contrast, while persistent-modes may affect mode awareness (i.e., the user’s ability to perceive and understand the system’s current operational state and the actions it affords), it could potentially be improved with XR's richer feedback channels, e.g., visual, haptic, and auditory~\cite{sellen1992prevention}. 
Prior work in XR individually studied switching techniques with the dominant/nondominant hand (DH / NDH)~\cite{surale2019experimental, park2020analysis},  eye gaze~\cite{pfeuffer2017gaze+, pfeuffer2020empirical}, or voice~\cite{shi2024experimental, chen2020disambiguation}. Options for subselection include gaze pointing with dwell~\cite{majaranta2009fast, hansen2018fitts},  pinch~\cite{pfeuffer2017gaze+, mutasim2021pinch}, and voice ~\cite{parisay2021eyetap, zhao2022eyesaycorrect}. To date, only one multi-selection work \cite{PinchCatcher} studied semi-pinch alongside gaze as a quasi-mode for a small number of targets (2-6 targets). Thus, it is unclear whether their results scale to a larger number of targets, which in turn would amplify mode-switching and therefore, multi-selection costs.

To explore multi-selection performance in XR, we ask the following research questions --- \textbf{RQ1:} \textit{How do controller-free quasi- and persistent-mode-switching strategies affect user performance for multi-selection in XR?} \textbf{RQ2:} \textit{How do pinch-based mode-switching techniques extend to a larger number of targets for multi-selection?} \textbf{RQ3:} \textit{How do different combinations of mode-switching and subselection techniques, such as gaze+pinch or hands-free methods like voice and dwell, influence multi-selection performance in XR?} To answer these, our user study compares four mode-switching techniques: two \qmode (\Fullpinch and \Semipinch) and two \pmode (\Doublepinch and \Voice), and three subselection techniques: Gaze+Dwell (\Dwell), Gaze+Pinch (\Pinch), and Gaze+Voice (\VSelect), to multi-select 6, 8, or 10 targets (\autoref{fig:teaser}).
Results revealed:

\begin{itemize}
    \item Participants switched mode faster with \qmode-mode techniques. Yet, \pmode-mode, especially \Doublepinch, consistently outperformed other switching techniques in terms of task completion time, error rate, and efficiency.
    
    \item \Doublepinch paired with \Pinch resulted in the best task completion time, error rate, and overall efficiency.

    \item \Semipinch produced the highest mode and selection error rate, and the lowest efficiency. Moreover, it was rated as most fatiguing, least usable, and showed the highest decline in efficiency with increasing targets.
    
    \item 17 and 15 of 30 participants preferred \Doublepinch mode-switching and \Dwell subselection technique, respectively.
\end{itemize}

Our contributions include: (1) a systematic empirical analysis of all the combinations of four mode-switching and three subselection techniques, informing design recommendations for gaze-based multi-selection of moderate to large target sets in XR, (2)~showing that \pmode-\Doublepinch is the most effective among the four investigated mode-switching techniques, and \Doublepinch~with~\Pinch outperformed other alternatives, and (3) demonstrating that \qmode-\Semipinch is \textit{not} a viable mode-switching option.

\section{Related Work}

Multi-selection can follow either a \textit{serial} or a \textit{parallel} workflow. Serial selection incrementally confirms objects one-by-one~\cite{wills1996selection, laviola20173d, lucas2005design, wu2023point}, while parallel selection defines a region (e.g., lasso) that selects multiple objects at once \cite{shi2021exploring, wu2023point, zhang2023multi}. Regardless of approach, users must first switch from single- to multi-selection mode, making mode-switching a core component of multi-selection.

\subsection{Mode-Switching in XR}

With controllers, mode-switching has been studied via buttons or menus~\cite{fernandez2022multi, wu2023point}, and is nowadays established in applications~\cite{shapelab_site_online, mennuti_blocks_next_level_2018, google_tiltbrush_tools_online}. For example, ShapeLab requires pressing a menu button to open a panel, selecting a multi-selection tool (brush, marquee, or ring), and then performing grouping with ray-based pointing and controller trigger input. Aside from controllers, different hands-free alternatives have also been studied, for instance, head gestures and variations such as nodding~\cite{Zhao2017RealtimeHG}, translation, and roll~\cite{shi2021exploring}. Yet, distinguishing intentional from natural head movements is challenging~\cite{hou2023classifying}, camera-coupled rotations increase cybersickness risk~\cite{palmisano2020cybersickness}, and repeated/exaggerated motions cause physical strain~\cite{souchet2023narrative}. 
Voice is also considered a promising hands-free approach as it reduces physical gestures, fatigue, and offers ergonomic benefits~\cite{Oh2021A3H}. While recognition latency remains a challenge, recent studies report acceptable recognition accuracy for typical XR tasks~\cite{Oh2021A3H, Jayasri2023ABR, valladares2025device, javaheri2025llms}, which motivated us to include it in our study. 

Hand gestures are intuitive and a prime candidate for mode-switching. Surale et al.~\cite{surale2019experimental} compared DH vs. NDH, finding NDH faster and more accurate. Smith et al.~\cite{smith2019experimental} reported similar findings, with follow-up work confirming scalability across tasks~\cite{Smith2020EvaluatingTS}. These results suggest that assigning mode-switch to the NDH improves performance, freeing up the DH for precise actions. Thus, in our work, we used the NDH for mode-switching. However, prior work also shows that complex or sustained mid-air gestures can lead to fatigue and interfere with fine manipulation~\cite{schon2023tailor}. In contrast, minimal gestures, such as a pinch, remain practical and efficient, motivating us to study this gesture further.

For 2D user interfaces (UIs), Pfeuffer et al.~\cite{Pfeuffer2015GazeShiftingDI} explored gaze with pen interactions and demonstrated use cases for a potential multimodal mode-switching mechanism. Fleischhauer et al.~\cite{hu2023gaze} found that gaze enables efficient, low-effort switching, and combining gaze fixation with a manual trigger improves robustness. In XR, Kim et al.~\cite{PinchCatcher} adopted a similar multimodal approach within the gaze+pinch paradigm, introducing \textit{semi-pinch}, a \qmode-mode multi-selection technique for small sets of targets (2, 4, and 6). They defined semi-pinch as a state ``on the way to'' a full-pinch, with final confirmation or end of multi-selection triggered by a full-pinch gesture. They compared five subselection techniques and found no significant differences in completion time. Nonetheless, semi-pinch combined with Swipe or Dwell was most preferred, and semi-pinch and Swipe yielded fewer errors. Yet, semi-pinch was not compared with other mode-switching techniques. While a subset of our study directly builds on their work~\cite{PinchCatcher}, here we also investigate semi-pinch alongside additional mode-switching techniques and evaluate its behavior when scaled to a larger number of targets.

\subsection{(Sub)Selection Techniques for Gaze-Based Systems}

In point-select tasks, users instinctively direct their gaze toward objects of interest~\cite{majaranta2014eye, liebling2014gaze, piumsomboon2017exploring, mutasim2021pinch}. Although gaze is fast~\cite{bahill1975main}, it lacks an explicit confirmation signal, leading to the \textit{Midas Touch} problem, where mere looking may trigger unintended selections~\cite{jacob1990you, penkar2012designing, isomoto2022interaction}. Thus, researchers have explored several selection techniques to complement interaction with gaze, e.g., ~\cite{mutasim2021pinch, pai2019assessing, esteves2020comparing, chatterjee2015gaze+, lu2020blinks}. However, \textit{dwell} still remains the most common solution to date~\cite{mutasim2025there}, where a target is confirmed by maintaining gaze on it for a fixed duration~\cite{majaranta2009fast, hansen2018fitts}. While a range of dwell durations has been explored~\cite{yi2022gazedock, mutasim2021pinch, mutasim2025there}, dwell still suffers from issues like latency and error~\cite{Patidar2014, Sidenmark2019_2, PinchCatcher}.

In contrast, hand gestures relatively better complement gaze as they provide more expressive and deliberate control that balances gaze’s rapid but implicit pointing~\cite{bragdon2011gesture, hyrskykari2012gaze}. This multimodal combination, where the eyes target and the hands confirm, has demonstrated strong potential across a wide range of interaction paradigms~\cite{jacob1990you, zhai1999manual, pfeuffer2014gaze, Pfeuffer2015GazeShiftingDI, chatterjee2015gaze+, mutasim2021pinch}. Gaze and hand now support tasks such as one-handed menu control~\cite{pfeuffer2023palmgazer}, 3D sketching~\cite{bashar2025depth3dsketch}, and accessibility~\cite{sidenmark2023vergence}, with the \textit{gaze+pinch} model emerging as a key interaction paradigm within this design space~\cite{pfeuffer2017gaze+, pfeuffer2024design, mutasim2021pinch}.

Voice has also been explored as a confirmation modality. Unlike pinch, voice is hands-free, eliminates mid-air arm fatigue, and other related issues like gesture ambiguity and hand-tracking noise~\cite{monteiro2025beyond}. Prior work showed that voice can be used to confirm selections~\cite{piumsomboon2017exploring, parisay2021eyetap, zhao2022eyesaycorrect}, issue commands in gaze-directed interaction~\cite{khan2021gavin}, or enhance accessibility for users with limited motor control~\cite{zhao2022eyesaycorrect, dondi2023gaze}. These advantages make voice a practical option for subselection in XR.

\section{Design and Implementation of Multi-Selection Techniques}

\subsection{Rationale for Chosen Techniques}

\subsubsection{Multi-Selection Techniques}

Our goal is \textit{not} to propose novel solutions, but to systematically evaluate representative controller-free mode-switching and subselection strategies that are supported and widely adopted in current commercial XR platforms for gaze-based multi-selection. To identify suitable options, we reviewed interaction literature across barehand XR and multimodal selection, and focused on methods that (1) require no external devices or UI widgets, (2) are already supported in commercial headsets (Meta Quest, HoloLens, Vision Pro), (3) use natural, low-learning gestures, and (4) can be executed continuously during mid-air interactions.

Thus, we selected \Fullpinch and \Semipinch, representing the \qmode forms of mid-air clutching~\cite{pfeuffer2017gaze+, PinchCatcher}, where a continuous gesture maintains the mode. \Doublepinch is chosen as a \pmode alternative because double-tap and double-pinch are widely used, easy to learn, and can be reliably detected~\cite{10.1145/1842993.1842997, fashimpaur2023investigating}. Although not a sensorimotor equivalent to pinch-based approaches, \Voice is chosen for pragmatic reasons as it is now native on major commercial virtual reality (VR) headsets and augmented reality (AR) glasses, and provides a convenient hands-free modality for \pmode option~\cite{parisay2021eyetap, zhao2022eyesaycorrect}. We excluded techniques that rely on digital menus~\cite{pfeuffer2020empirical}, surface contact~\cite{pfeuffer2014gaze, Pfeuffer2015GazeShiftingDI, hu2023gaze}, wearable devices~\cite{yeo2019wrist, lang2023multimodal}, or micro-gestures~\cite{shi2024experimental}, as these either require additional hardware, introduce fatigue, or impose high learning and memorization costs that make them less suitable for sustained freehand interaction.

For subselection, we selected \textit{Dwell}, \textit{Pinch}, and \textit{Voice} as they represent the three dominant confirmation modalities in gaze-based XR: gaze-only~\cite{majaranta2009fast}, hand-gesture~\cite{mutasim2021pinch}, and voice commands~\cite{zhao2022eyesaycorrect}. These three cover distinct trade-offs in effort, speed, and error control, and are directly comparable as all three complement gaze pointing well.

\subsubsection{Serial Multi-Selection}

Parallel multi-selection techniques perform best when desired targets are (densely) clustered. However, such clustering is not representative of general multi-selection scenarios, where targets are typically scattered and/or surrounded by distractors, e.g., selecting files/images to upload to social media. In such cases, especially for a relatively moderate number of targets (e.g., $\leq 10$)~\cite{wu2023point}, serial methods are more suitable. Thus, we decided to investigate \emph{serial} grouping for moderate target counts to explore more realistic use cases.

\subsubsection{Bimanual Multi-Selection}

Instead of introducing unfamiliar gesture vocabularies, we chose pinch-based interactions because they are already widely used in XR systems and require minimal learning effort. For example, a single pinch with the DH naturally maps to single-object selection, while a bimanual pinch is often used for transformations such as resizing or rotating~\cite{hinrichs2011gestures, chaconas2018evaluation}. Replacing pinch with new gestures would increase the learning burden by forcing users to learn an additional gesture vocabulary~\cite{shi2024experimental, PinchCatcher}. Instead, we adopt a bimanual interaction approach that complements gaze --- the NDH defines spatial context (e.g., mode, frame, scale), while the DH executes precise operations such as selection, translation, or rotation. This design is grounded in Guiard’s~\cite{guiard1987asymmetric} kinematic-chain model of asymmetric hand roles, which has repeatedly shown efficiency and fluidity gains in compound Human-Computer Interaction (HCI) tasks. Accordingly, we place mode-switching actions in the NDH and familiar pinch gestures on the DH for subselection. 

\subsection{Mode-Switching Techniques}
Among the four chosen techniques, three are based on the gaze+ pinch interaction paradigm~\cite{pfeuffer2024design} (\Fullpinch, \Semipinch, and \Doublepinch) while one leverages a hands-free modality (\Voice). Following prior work~\cite{surale2019experimental, smith2019experimental, Smith2020EvaluatingTS}, we assigned all the pinch variants to the NDH. More details about all the techniques are provided below: 

\textit{\textbf{\Fullpinch.}}
This technique requires the user to maintain a full-pinch, i.e., pinch and hold, to activate and sustain multi-selection mode. A pinch is detected when the thumb and index fingertips' distance is $< 2~cm$ (\autoref{fig:teaser}). When the pinch is released (or the system loses recognition of the pinch gesture), i.e., when the two fingertips' distance is $> 7~cm$, the system waits $250~ms$, before deactivating the mode and finalizing the group. The $250~ms$ gap was chosen based on the touchscreen multi-selection touch-dwell approach~\cite{appleSelectItems} to prevent errors due to tracking issues.

\textit{\textbf{\Semipinch.}}
Following Kim et al.’s design~\cite{PinchCatcher}, multi-selection is activated when the distance between thumb and index fingertips is within $2$–$7~cm$ (\autoref{fig:teaser}). The grouping is finalized with a full-pinch, after which the system waits $250~ms$ before ending the trial. Similar to existing, familiar, and realistic \qmode multi-selection behavior on 2D devices, where an accidental key release or momentary loss of touch input reverts the interaction to single-selection mode, our system also returns to single-selection when the semi-pinch gesture is (unintentionally) released (i.e., $> 7~cm$). Matching established conventions, we also designed the single-selection mode to revert any previously selected object(s) to their default state upon each new selection. Importantly, however, the trial itself remains open and is not terminated. Further, to minimize the effects of unintended mode changes, we provided color-coded visual feedback indicating the current selection mode~(see \autoref{fig:grid} and Section \ref{sec:exp_task}) throughout the interaction. This design choice somewhat enables users to notice and recover from silent mode drops.

\textit{\textbf{\Doublepinch.}}
A double-pinch~\cite{fashimpaur2023investigating} is recognized when a full-pinch is released and performed again within $350~ms$ of the first pinch (\autoref{fig:teaser}). Another double-pinch deactivates the mode, thus finalizing the selected group. The $350~ms$ interval was chosen based on pilot studies and prior work that suggest a range of $300$–$400~ms$~\cite{10.1145/1842993.1842997, nishida2023single} to balance speed and reliability. During pilots, shorter intervals led to missed activations, while longer ones increased false positives.

\textit{\textbf{\Voice.}}
Following prior work~\cite{schmitt2021voice}, the vocabulary was chosen to be short, unambiguous, and semantically distinct to minimize voice recognition errors. Participants could choose from a small set of activation words (\textit{group}, \textit{multi}, \textit{batch}, \textit{multimode}) to activate multi-selection, and distinct terms (\textit{complete}, \textit{done}, \textit{single}, \textit{finish}) to deactivate and finalize the group. Enabling users to select the phrase most natural to them could potentially reduce cognitive load, while ensuring commands were concise and disjoint enough to avoid misrecognition. Longer or ambiguous utterances (e.g., ``switch'') were deliberately excluded, as they have been shown to increase recognition latency and frustration~\cite{wang2025exploring}. 

\textit{Other Techniques Considered.}
A \textit{single NDH pinch (and release)} could act as a \pmode toggle, but pilot testing revealed that users often got confused and struggled to distinguish NDH pinch for mode-switching from DH pinch for subselection. This confusion was most prominent at mode exit, which increased errors, cognitive load, and task completion time. Another candidate was a \textit{\qmode voice mode} (sustained vocalization), such as humming \cite{hedeshy2021hummer}, which was also discarded as it is fatiguing~\cite{schmitt2021voice}. We also excluded \textit{single modality switching} via \textit{menu UIs}~\cite{shapelab_site_online, mennuti_blocks_next_level_2018, google_tiltbrush_tools_online}, since single modalities disrupt the natural eye–hand division of labor~\cite{pfeuffer2017gaze+, hirzle2022understanding}.

\subsection{Subselection Techniques}

\hspace{\parindent}\textit{\textbf{Gaze+Dwell (\Dwell).}}
To select an object, the user's gaze cursor has to maintain fixation on the object for $450~ms$~\cite{penkar2012designing, hansen2018fitts}. The dwell duration was chosen based on a pilot and prior work~\cite{mutasim2025there, PinchCatcher, Rajanna2018, Mutasim2025GEM_Gaze}.

\textit{\textbf{Gaze+Pinch (\Pinch).}}
A full-pinch with the DH selects the object that the user's gaze cursor is on \cite{pfeuffer2017gaze+, mutasim2021pinch}. 

\textit{\textbf{Gaze+Voice (\VSelect).}}
The user gazes at the target and then issues a spoken command to select it. As above, we again used a small set of unique commands (\textit{select}, \textit{add}, \textit{pick}, \textit{choose}, \textit{take}).

\section{User Study}

To assess the performance of the multi-selection techniques, and following established XR study guidelines ~\cite{bergstrom2021evaluate}, we designed a user study within a scattered serial target grouping task~\cite{lucas2005design, wu2023point, PinchCatcher}. We did not use Fitts’ law~\cite{hansen2018fitts, bashar2025effect} or other single-target pointing tasks (e.g., ~\cite{mutasim2022saccades, Choe2019}), since our objective was to evaluate naturalistic multi-selection performance and not isolated pointing. To avoid confounds such as depth perception, occlusion, and spatial search, targets were arranged in a 2D grid placed within a 3D VR environment, allowing performance differences to be primarily attributed directly to mode-switching and subselection. This controlled layout is consistent with prior multi-selection research~\cite{lucas2005design, wu2023point, shi2024experimental}, which typically investigates simplified spatial arrangements and the core interaction mechanism first before moving on to complex scenarios. 

\begin{figure}[ht!]
    \centering
    \includegraphics[width=1\linewidth]{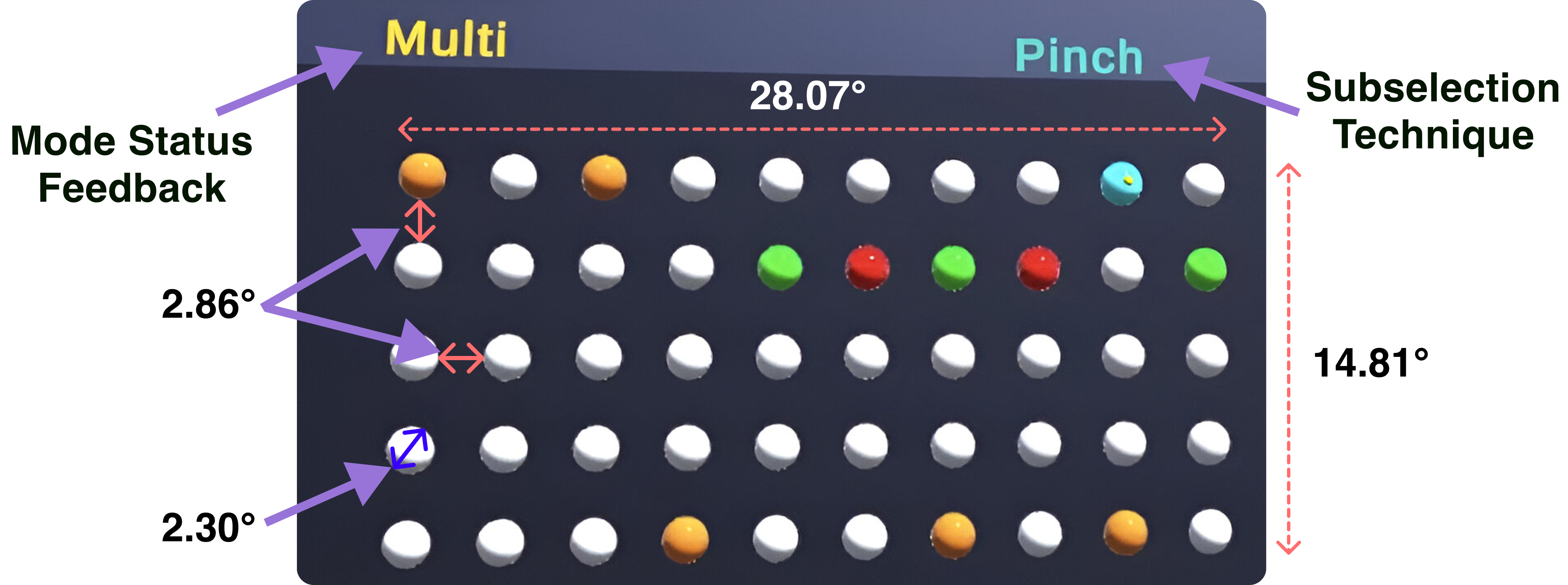}
    \caption{The experimental task with the mode (top-left) and subselection (top-right) indicators.}
    \Description{Overall Experimental Environment and Stimulus Design. The experiment employed a flat grid layout for object presentation. A total of 50 spherical objects of size 2.30 degrees were evenly distributed across the 28.07 degrees by 14.81 degrees grid with a uniform spacing of 2.86 degrees between them. When participants fixated their gaze on an object, its color changed to teal, indicating that the object was being looked at. Target objects were initially highlighted in orange; upon successful confirmation, they turned green, whereas incorrect selections turned red. A mode-status indicator was displayed in the top-left corner of the grid: when the multi-selection mode was active, the text ``Multi'' appeared in yellow, while in single-selection mode, ``Single'' appeared in white. In the top-right corner, the currently active subselection technique was displayed in teal.}
    \label{fig:grid}
\end{figure}

For similar reasons, we also did not investigate deselection, i.e., reselecting an already selected object to remove it from the current group. We argue that deselection is the same motor interaction as selection, and is thus the same underlying mechanism rather than a distinct operation in serial multi-selection. Moreover, as we designed our task to visually show correctly (green) and incorrectly (red) selected objects (see \autoref{fig:grid}), we somewhat reduced the cognitive confound involved in identifying the need for deselection. Also, we did not deem it suitable to explore scrolling or navigation actions, as they constitute fundamentally different interaction processes.
Still, to preserve ecological variability while maintaining experimental repeatability, we randomize target positions across trials similar to prior work~\cite{bergstrom2021evaluate, lucas2005design, wu2023point}.

\subsection{Experimental Task}\label{sec:exp_task}

As serial multi-selection is more suitable for selecting small to moderate number of targets, we chose to present 6, 8, or 10 orange targets in a single trial. The rest of the objects in the $5\times10$ flat rectangular grid were colored in white and acted as distractors. To provide visual feedback, when a participant’s gaze landed on an object, it turned teal. Correctly selected targets changed to green, while incorrect selections changed the distractors to red. The current selection mode was shown in the top-left corner of the grid and was color-coded: ``Multi'' selection appeared in yellow, while ``Single'' appeared in white. Also, a subselection indicator was shown in the top-right corner as visual feedback  (see \autoref{fig:grid}).

Each trial began with an inter-trial interval scene containing a ``Start'' button at arm’s reach, which participants tapped with their finger to start the trial. Participants were then instructed to group all orange targets while avoiding distractors, and explicitly told to prioritize accuracy over speed. This design choice parallels real-world tasks, where the cost of incomplete or incorrect grouping outweighs that of minor delays.

The grid was placed $10~m$ in front of the participants. Objects were evenly spaced with a edge-to-edge gap of $2.86^{\circ}$, and the entire grid was $28.07^{\circ} \times 14.81^{\circ}$. Each sphere had a diameter of $2.30^{\circ}$; however, to compensate for minor inaccuracies in eye tracking, invisible colliders were enlarged 1.5 times, i.e., to $3.45^{\circ}$, following prior work~\cite{lystbaek2024hands, PinchCatcher}.

\subsection{Experimental Design}\label{sec:exp_design}

We implemented a within-subjects design with three independent variables --- \textbf{mode-switching} (\Doublepinch, \Fullpinch, \Semipinch, and \Voice), \textbf{subselection} (\Dwell, \Pinch, and \VSelect), and, to vary task difficulty and enable analysis of scalability effects, \textbf{number of targets} (6, 8, or 10). Each participant completed all combinations, with three trial repetitions per condition, resulting in a total of $4_{mode-switching} \times 3_{subselection} \times 3_{\#targets} \times 3_{repetitions} = 108$ trials per participant. To mitigate order effects, the conditions were counterbalanced using a Latin Square design.

\subsection{Apparatus}

The study was implemented in Unity (6000.0.45f1) for the Meta Quest Pro VR headset, which features $90~Hz$ refresh rate, 106$^{\circ}$ horizontal field-of-view, integrated with eye trackers sampling at $72~ Hz$~\cite{rocchi2025comparison}. Gaze and hand-based interactions were implemented using the Meta XR All-in-One SDK (v74.0.1) and were smoothed using a 1€ Filter~\cite{casiez20121}. Pinch was detected using the Meta Hand Pose Detection SDK, which also tracks fingertip distances. For voice recognition, we used the open-source Vosk speech recognition toolkit\footnote{Vosk speech recognition toolkit: \url{https://github.com/alphacep/vosk-api}} (Kaldi-based)~\cite{owen2025improving}, integrated into Unity via the Meta XR Voice SDK for microphone access. The toolkit was run locally on-device to reduce recognition latency.

We measured interaction latency for both pinch and voice input in early pilots. End-to-end pinch detection latency was quantified using high-speed video recording ($\geq$ 240~fps) of participants performing pinch gestures while an immediate visual response was triggered by the system upon detection. Latency was computed as the difference between the first video frame capturing the physical thumb–index contact (gesture onset) and the first frame showing the corresponding system response. Using this method, pinch events (full-pinch and release) were recognized within 30-50 $ms$ from gesture onset, matching previous work~\cite{godden2025robotic}. 

Voice latency of about 120-150 $ms$ was measured from speech onset to the corresponding system response~\cite{xie2024intelligent}. Speech onset was identified as the first audio frame exceeding the recognizer's voice-activity detection threshold, and system response was defined as the frame in which the command triggered a visible state change.

\subsection{Participants}

We recruited 30 participants for the study (16 female, 14 male; $M_{Age}$ = 26.4, $SD_{Age}$ = 2.40). 26 were right-handed, and 4 were left-handed. 9 wore glasses, 2 used contact lenses, and 2 had undergone vision correction surgery. Participants came from varied backgrounds, including engineering, advertising, and customer service, providing a broad range of perspectives. They self-rated their familiarity with AR/VR/XR on a 6-point Likert scale (0 = no experience, 5 = expert), reporting moderate overall experience ($M$ = 3.50, $SD$ = 1.04). By modality, experience was highest with controllers ($M$ = 3.86, $SD$ = 1.07), followed by hand interactions ($M$ = 2.30, $SD$ = 0.95), gaze ($M$ = 1.96, $SD$ = 1.79), and voice ($M$ = 1.03, $SD$ = 0.92). The study protocol was approved by the institution’s ethics board.

\subsection{Procedure}

We first obtained consent, collected demographic information, and then briefed participants on the study tasks. Then they put on the VR headset and completed the built-in eye-tracking calibration, which was repeated whenever the headset was re-worn to ensure accuracy \cite{mutasim2022saccades}.

Each condition had three phases. In the \textit{training phase}, participants practiced the mode-switching + subselection technique for three one-minute trials with only 10 spheres in the grid and no specified targets. In total, 36 training trials were conducted, which were excluded from the analysis. In the \textit{experimental phase}, participants performed the randomized target grouping task. They were instructed to complete the three trial repetitions of each condition continuously. In the \textit{questionnaire phase}, participants filled out questionnaires evaluating their experience with that specific condition, and were offered short breaks before the next one.  

After completing all conditions, participants ranked the mode-switching and subselection techniques in order of preference and took part in a short semi-structured interview. During the interview, participants were asked to elaborate on the rationale behind their rankings and reflect on their experience. Overall, the study lasted about an hour.

\begin{figure*}[htbp!]
    \centering
    \includegraphics[width=1\linewidth]{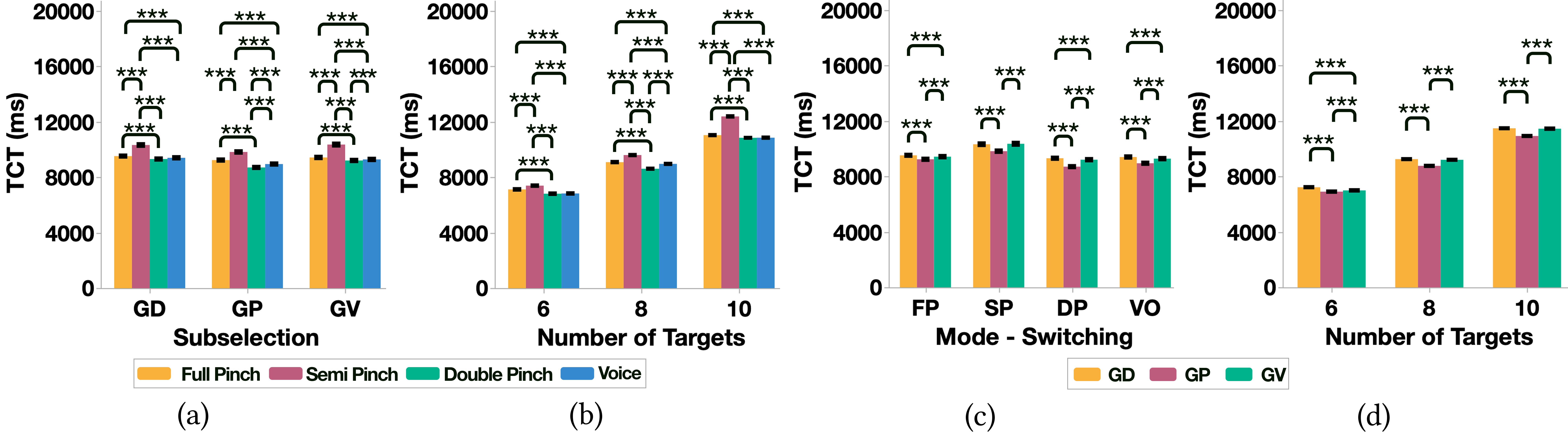}

    \caption{Task completion time (TCT) results for the interaction between (a) mode-switching and subselection, (b) mode-switching and number of targets, (c) subselection and mode-switching, and (d) subselection and number of targets.}   
    \label{fig:TCT}
    \Description{Four grouped bar charts show task completion time (TCT, in milliseconds). (a) For each subselection technique, bars represent switching techniques Full-Pinch (FP), Semi-Pinch (SP), Double-Pinch (DP), and Voice (VO); SP consistently has the longest TCT, while DP and VO are fastest. (b) For each mode-switching technique, TCT increases as the number of targets rises from 6 to 10, with SP remaining slowest overall. (c) Switching techniques are compared across subselection methods, showing lower TCT for DP and VO than for FP and SP. (d) Across subselection techniques, TCT increases with the number of targets. Error bars indicate 95\% confidence intervals, and asterisks mark statistically significant differences.}
\end{figure*}

\subsection{Evaluation Metrics} \label{sec:eval}
\begin{itemize}
    \item \textit{Task Completion Time (TCT)}: The interval between tapping the ``Start'' button and performing the action that ended the trial.
    
    \item \textit{Mode-Switching Time (MST)}: Time from tapping the ``Start'' button to the first successful mode change. MST was calculated twice per trial: once each for entering and exiting multi-selection mode. For example, with \Doublepinch, the first MST was measured from the press of the ``Start'' button until the system registered the second pinch, which triggered multi-selection mode. The second MST was measured from the moment the last sphere was selected until the next double-pinch was detected, finalizing the group and returning to single-selection mode.

    \item \textit{Mode Error (ME)}: 
    The sum of how many times (1) participants attempted multi-selection without first switching modes~\cite{hu2023gaze}, and (2) the system reverted to single-selection during multi-mode.
    
    \item \textit{Accidental Subselection Ratio (ASR)}: Number of distractors (white spheres in \autoref{fig:grid}) unintentionally grouped during a trial, divided by the total number of subselections in that trial~\cite{PinchCatcher}.
    
    \item \textit{Error Rate (ER)}: 
    Defined as the sum of the number of missed targets and the number of distractors selected, divided by the total number of grouped objects~\cite{PinchCatcher, shi2024experimental}.
    
    \item \textit{Inverse Efficiency (IE)}: 
    We calculated the success rate as the percentage of error-free trials ~\cite{statsenko2020applying, PinchCatcher} and computed IE as TCT divided by success rate, capturing the speed–accuracy trade-off. Higher values indicate lower efficiency.

    \item \textit{Number of Actions (NoA)}: Total count of discrete actions per trial, including both mode-switching gestures/commands and subselections~\cite{shi2024experimental}.

    \item \textit{Preference and Subjective Task Load}: After each condition, participants completed the System Usability Scale (SUS)~\cite{brooke2013sus}, NASA-RTLX~\cite{hart2006nasa}, and Borg CR10~\cite{wentzel2020improving}. Post-study, participants ranked the techniques and elaborated on their choices in a semi-structured interview.    
\end{itemize}

\section{Results}

Across all conditions and participants, we recorded 27864 multi-selections (including accidental subselections). Data was analyzed using three-way repeated measures (RM) ANOVA (see Section \ref{sec:exp_design}) and Bonferroni correction for post-hoc analysis. We considered continuous variables to be normally distributed when skewness and kurtosis were within $\pm1$~\cite{Hair2014multivariate, Mallery2003spss}. For non-(log)-normal variables, we used the Wilcoxon Signed-Rank test. In cases where sphericity was violated, Greenhouse-Geisser corrections were applied. 

Trials in which participants skipped or failed to complete the task were excluded from the analysis. In addition, outlier trials with completion times greater than three standard deviations above the mean were removed (22 trials, 2.55\%), as well as two accidentally skipped trials (0.25\%), totaling 24 trials (2.80\%) being excluded. For questionnaire data (Borg CR10, SUS, NASA-RTLX), we used the Friedman test with the Wilcoxon Signed-Rank test for post-hoc analysis (Bonferroni corrected). For brevity, we only report the most relevant significant results here. Complete RM ANOVA results and analysis of the evaluation metric \textit{Hand Movement (HM) and Rotation (HR)} are provided in the supplemental materials. In all graphs, error bars represent 95\% confidence intervals, and statistical significance is annotated as: $*$ ($p < .05$), $**$ ($p < .01$), and $***$ ($p < .001$).

\subsection{Task Completion Time (TCT) -- \autoref{fig:TCT}}

A significant three-way interaction was found for TCT, $F(12,348) = 44.911, p < .001, \eta_p^2 = .608$. In addition, there was a significant interaction between mode-switching and subselection techniques, $F(6,174) = 11.515, p < .001, \eta_p^2 = .284$. Overall, participants completed the task fastest with \Doublepinch and slowest with \Semipinch (\autoref{fig:TCT}a). The interaction between mode-switching and number of targets was also significant, $F(6,174) = 162.52, \, p < .001, \, \eta_p^2 = .849$. Participants took more time to complete the task as the number of targets increased, with the sharpest rise for \Semipinch (\autoref{fig:TCT}b). \Pinch was the fastest, followed by \VSelect, while \Dwell was slowest~(\autoref{fig:TCT}c). There was also a significant interaction between subselection and the number of targets. Post-hoc analysis again showed that \Pinch was fastest (\autoref{fig:TCT}d). Taken together, \Doublepinch + \Pinch achieved the lowest completion times, whereas \Semipinch + \Dwell produced the highest.

\begin{figure}[ht!]
    \centering
    \begin{subfigure}[t]{0.45\columnwidth}
        \includegraphics[width=1\linewidth]{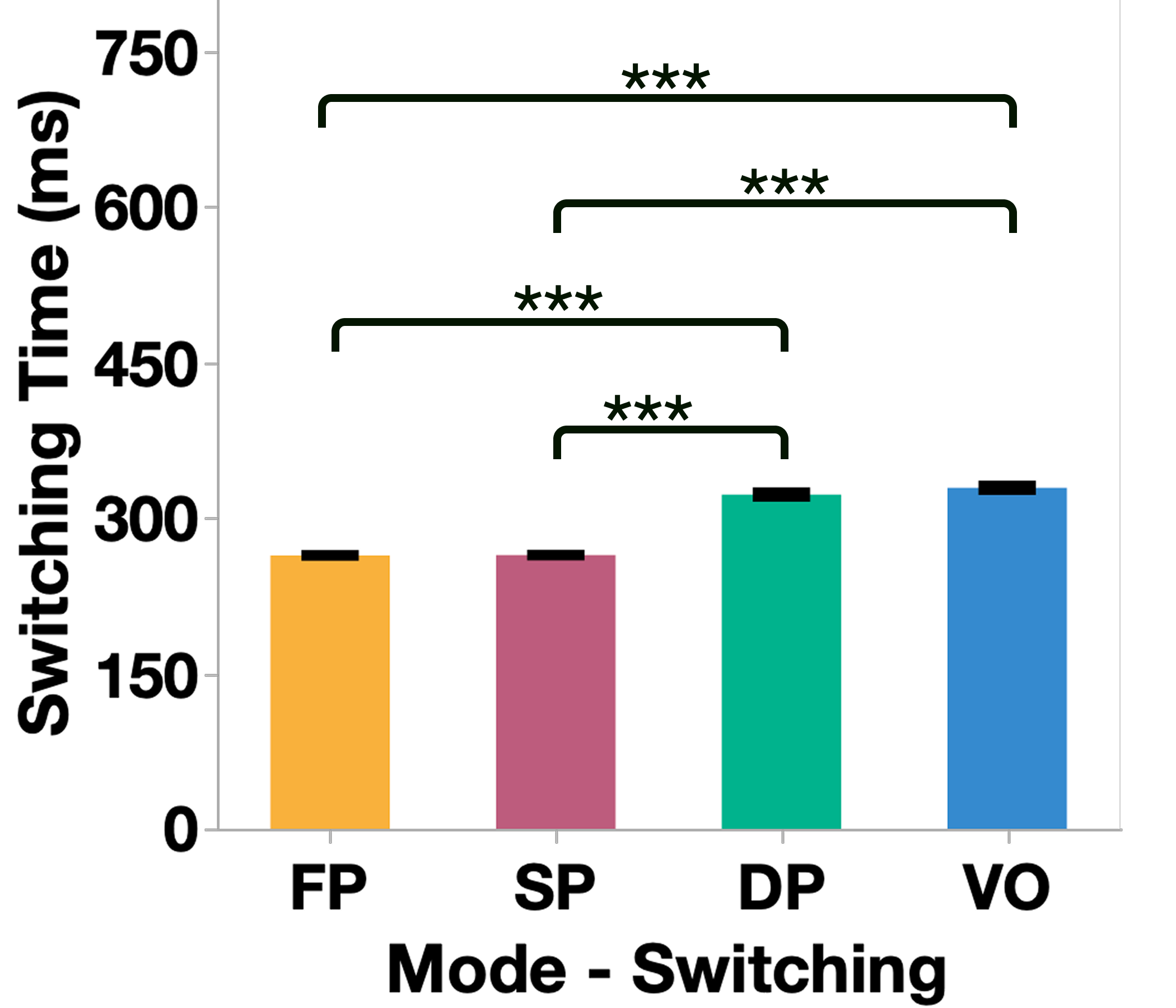}
        \caption{}
        \label{fig:mode-switch}        
    \end{subfigure}
    \hspace{0.05\columnwidth}
    \begin{subfigure}[t]{0.45\columnwidth}
        \includegraphics[width=1\linewidth]{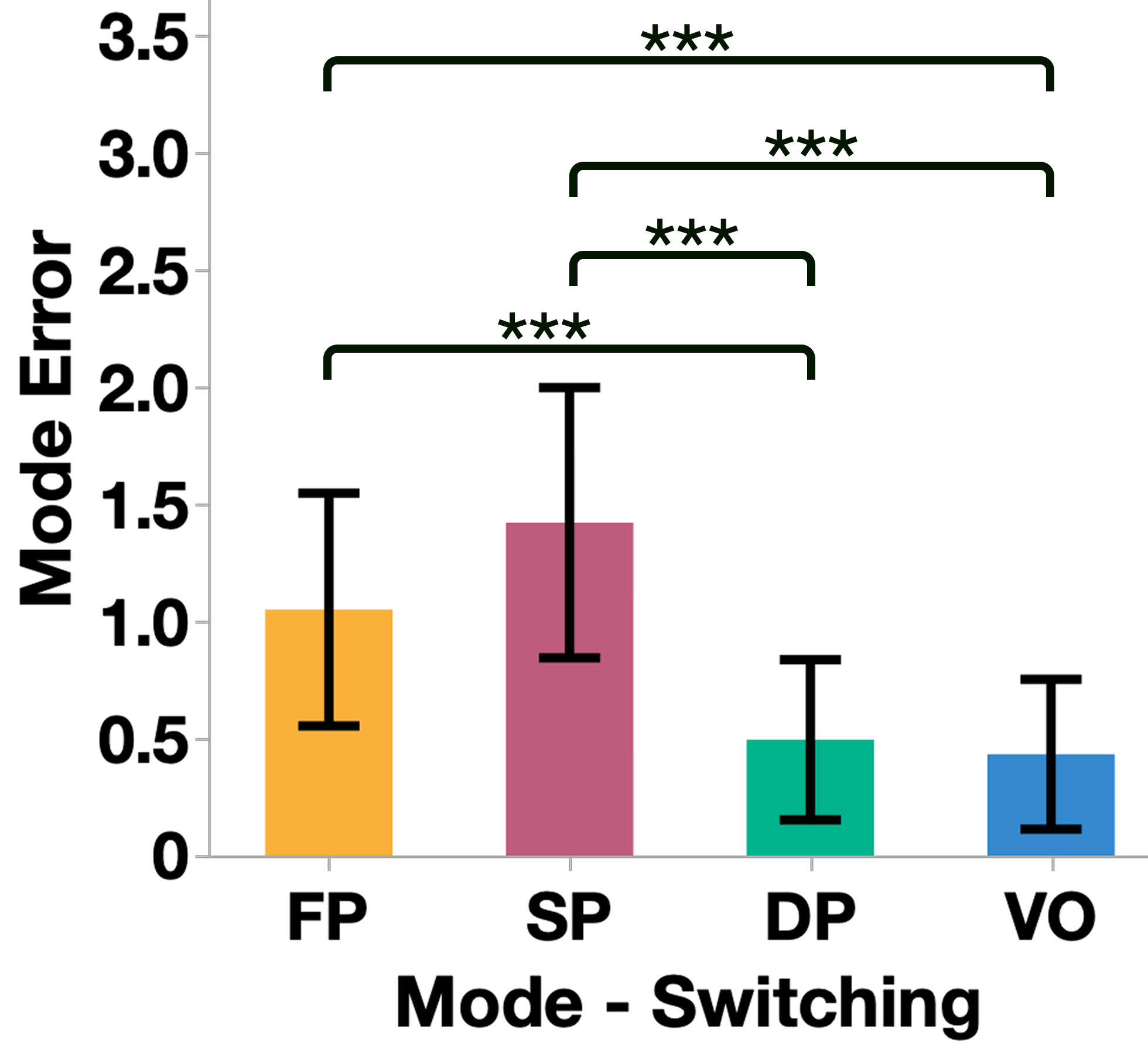}
        \caption{}
        \label{fig:mode-error}
    \end{subfigure}
    \caption{(a) Mode-switching time (MST) and (b) mode error (ME) results for the four mode-switching techniques.}   
    \label{fig:mode}
        \Description{The average mode-switching time (in milliseconds) and mode error for each mode-switching technique. (a) Switching time: FullPinch (FP) and SemiPinch (SP) are fastest and do not differ; DoublePinch (DP) and Voice (VO) are slower and comparable. (b) Mode error: SP exhibits the highest mode-related error rate, while FP falls in the middle, and DP and VO share the lowest rate. The error bars represent 95\% confidence intervals. Brackets above bars indicate post-hoc significance: * p < .05, ** p < .01, and *** p < .001.}
\end{figure}

\subsection{Mode-Switching Time (MST) -- \autoref{fig:mode-switch}}

The results showed that there was a significant main effect of mode-switching technique on MST, $F(1.672, 48.492) = 1339.69,  p < .001, \eta_p^2 = .98$. Post-hoc comparisons showed that both \Doublepinch and \Voice were significantly slower than \Fullpinch and \Semipinch, with no significant difference between either of the two pairs.

\subsection{Mode Error (ME) -- \autoref{fig:mode-error}} 

Only a significant main effect of the mode-switching technique was found, $F(3,87) = 6.516, p < .001, \eta_p^2 = .183$. \Semipinch produced the most errors, followed by \Fullpinch, whereas \Doublepinch and \Voice exhibited least errors. Post-hoc test confirmed that \Semipinch and \Fullpinch were significantly more error-prone than both \Doublepinch and \Voice.

\begin{figure*}[t!]
    \centering
    \includegraphics[width=0.75\linewidth]{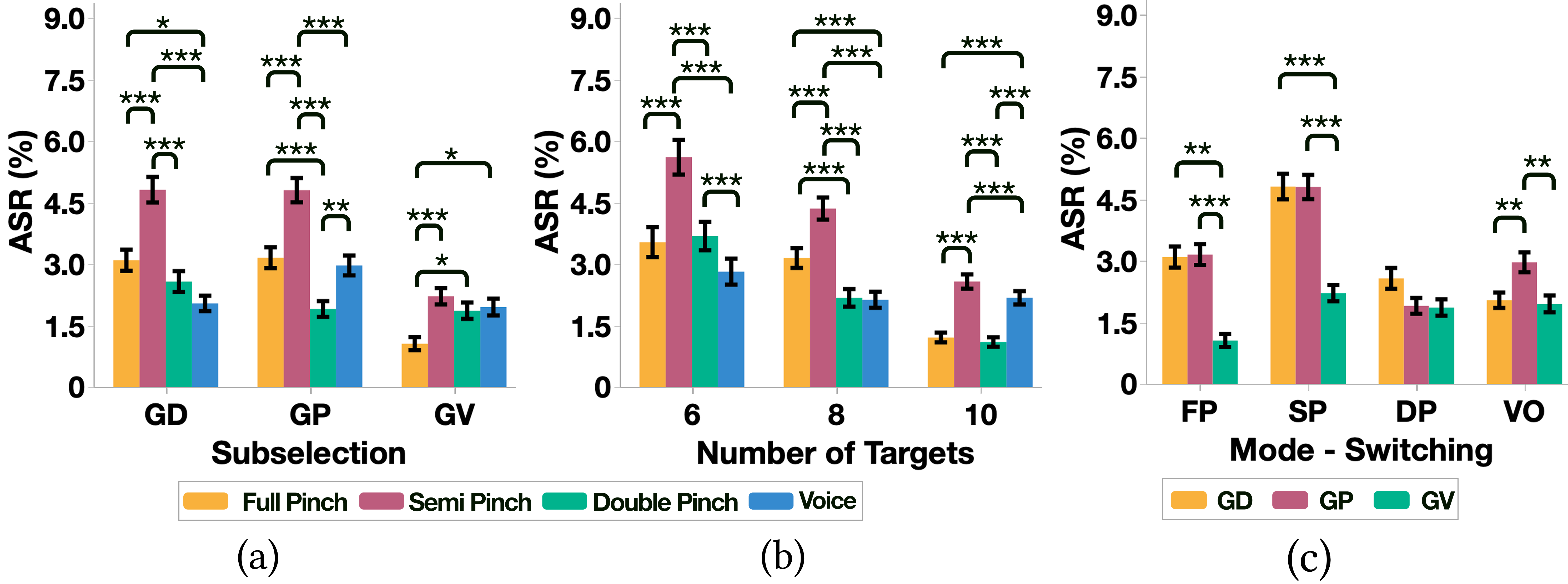}
    
    \caption{Accidental subselection ratio (ASR) results for the interaction between (a) mode-switching and subselection, (b) mode-switching and number of targets, and (c) subselection and mode-switching.}   
    \label{fig:ASR}
        \Description{Accidental Subselection Ratio (ASR) across conditions. (a) ASR by subselection technique (Dwell (GD), Pinch (GP), Voice (GV)) across all mode-switching techniques. GV produced higher ASR than GP and GD, while SemiPinch consistently yielded the highest values. (b) ASR by number of targets (6, 8, 10) across mode-switching techniques. ASR increased significantly with more targets, particularly for SemiPinch, whereas DoublePinch and Voice scaled more gracefully. (c) ASR by mode-switching technique (DoublePinch, FullPinch, SemiPinch, Voice) across subselection techniques. Semipinch produced significantly more accidental subselection than all other techniques, while DoublePinch and Voice yielded the lowest ASR. Quasi techniques, especially SemiPinch, were disproportionately affected by higher target counts, while persistent techniques maintained consistently low ASR. The error bars represent 95\% confidence intervals. Brackets above bars indicate post-hoc significance: * p < .05, ** p < .01, and *** p < .001.}
\end{figure*}

\begin{figure*}[t!]
    \centering
    \includegraphics[width=1\linewidth]{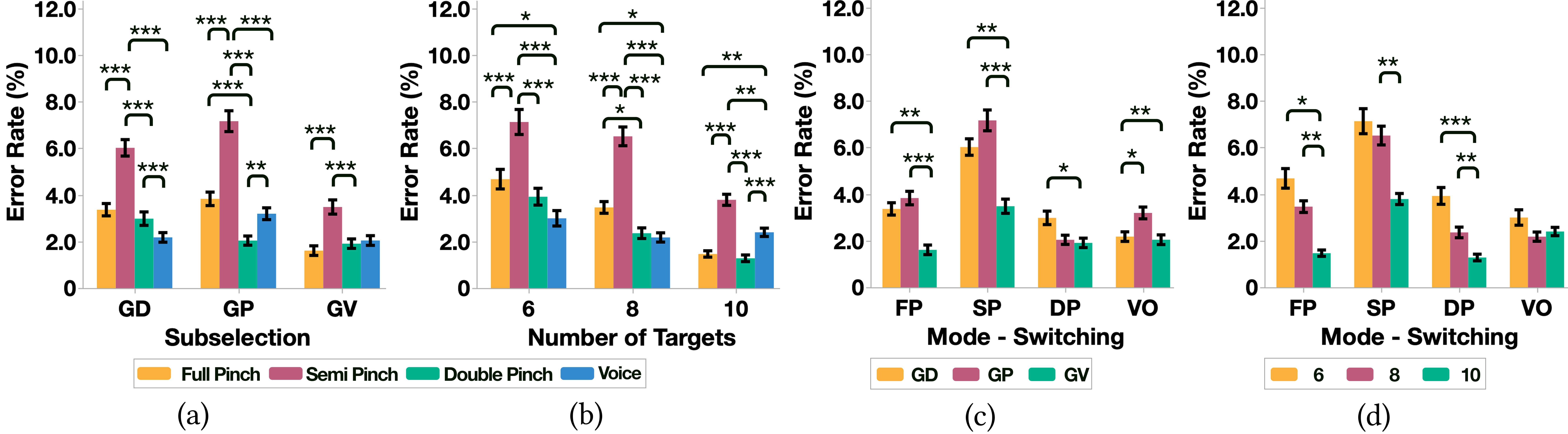}
    
    \caption{Error rate (ER) results for the interaction between (a) mode-switching and subselection, (b) mode-switching and number of targets, (c) subselection and mode-switching, and (d) number of targets and mode-switching.}   
    \label{fig:ER}
    \Description{Error Rate (ER) includes the number of ungrouped targets and distractors selected, divided by the total number of grouped objects per trial. The figure contains four grouped bar charts. (a) ER is shown by subselection techniques (Dwell (GD), Pinch (GP), Voice (GV)) across all mode-switching methods; GP generally shows the highest error rates, while GD and GV are lower. (b) ER is shown by mode-switching techniques (DoublePinch, FullPinch, SemiPinch, Voice) across a number of targets, with error rates increasing as the number of targets increases, particularly for SemiPinch. (c) ER across mode-switching across subselection techniques, showing higher errors for Semi Pinch than the other techniques. (d) The interaction between mode-switching and number of targets shows that error rates increase with larger target sets for all techniques. The error bars represent 95\% confidence intervals. Brackets above bars indicate post-hoc significance: * p < .05, ** p < .01, and *** p < .001.}
\end{figure*}

\subsection{Accidental Subselection Ratio (ASR) -- \autoref{fig:ASR}}

Significant interactions were observed between mode-switching and subselection techniques, $F(4.475,129.775) = 8.04, p < .001, \eta_p^2 = .217$, mode-switching and number of targets, $F(3.491,101.249) = 5.58, p < .001, \eta_p^2 = .161$, and a three-way interaction of all three factors, $F(7.336,212.750) = 2.38, p = .006, \eta_p^2 = .076$. Post-hoc tests showed that \Semipinch produced significantly higher ASR than all other switching techniques (\autoref{fig:ASR}a). For subselection, \Voice performed better in general (\autoref{fig:ASR}c). Participants selected fewer distractors when the number of targets increased, which was common across all mode-switching techniques (\autoref{fig:ASR}b). The three-way interaction indicated that \qmode techniques, particularly \Semipinch paired with \Pinch, increased ASR at lower target counts, whereas \pmode-modes maintained stability across subselection techniques and number of targets.

\subsection{Error Rate (ER) -- \autoref{fig:ER}}

Significant interactions were found between mode-switching and subselection techniques, $F(3.012,87.357) = 5.90, \, p < .001, \, \eta_p^2 = .169$, and between mode-switching and number of targets, $F(2.867,$ $83.129) = 4.22, \, p < .001, \, \eta_p^2 = .127$. Post-hoc revealed that \Semipinch produced the most errors (Figures \ref{fig:ER}a and \ref{fig:ER}b), while \VSelect exhibited the fewest errors (Figures \ref{fig:ER}c and \ref{fig:ER}d). Like ASR, ER showed a similar pattern where the errors decreased with increasing targets. 

\begin{figure*}[!hbt]
    \centering
    \includegraphics[width=1\linewidth]{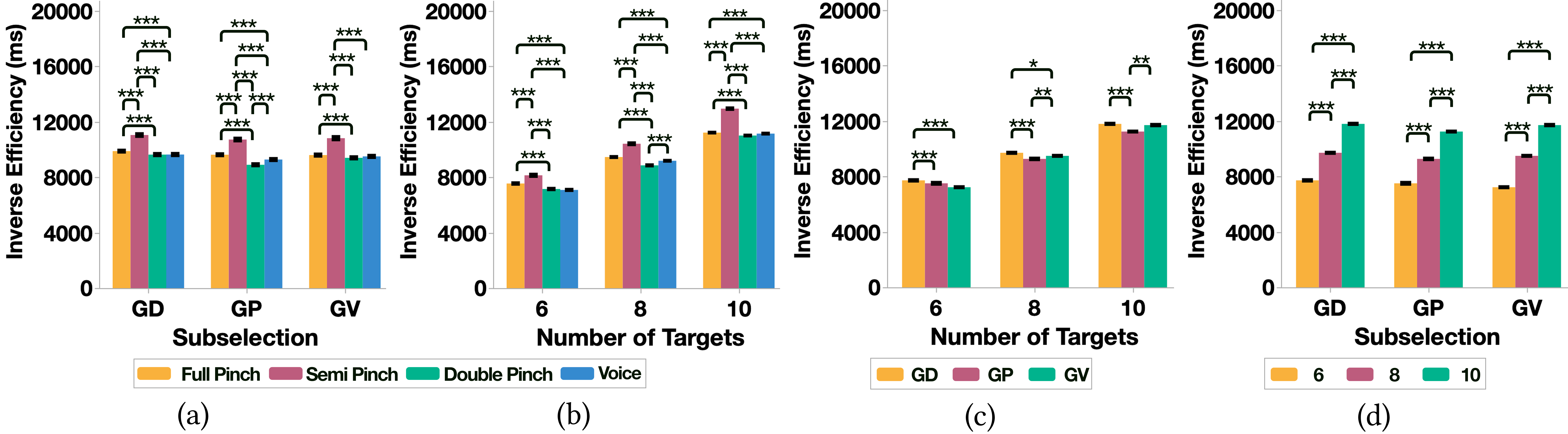}

    \caption{Inverse Efficiency (IE) results for the interaction between (a) mode-switching and subselection, (b) mode-switching and number of targets, (c) subselection and number of targets, and (d) number of targets and subselection.}   
    \label{fig:IE}
        \Description{Inverse Efficiency (IE) results across mode-switching techniques, subselection techniques, and number of targets. (a) IE by subselection technique, showing Pinch (GP) consistently outperforming Dwell (GD) and Voice (GV). (b) IE by mode-switching technique, where DoublePinch and Voice maintained lower IE compared to quasi techniques (FullPinch, SemiPinch). (c) Interaction of the subselection techniques with the number of targets. (d) Interaction of mode-switching technique with number of targets, highlighting that SemiPinch performed worse, showing steep increases in IE at higher target counts. The error bars represent 95\% confidence intervals. Brackets above bars indicate post-hoc significance: * p < .05, ** p < .01, and *** p < .001.}
\end{figure*}

\subsection{Inverse Efficiency (IE) -- \autoref{fig:IE}}

The IE data showed a significant two-way interaction between mode-switching and subselection, $F(6,174) = 10.78, \, p < .001, \, \eta_p^2 = .271$, mode-switching and number of targets, $F(3.880,112.506) = 13.84, \, p < .001, \, \eta_p^2 = .323$, and subselection and number of targets, $F(2.872,83.288) = 9.66, \, p < .001, \eta_p^2 = .250$. The three-way interaction was also significant, $F(6.204,179.918) = 11.05, \, p < .001, \, \eta_p^2 = .276$. Post-hoc revealed that \Semipinch and \Dwell (Figures \ref{fig:IE}a and \ref{fig:IE}d) resulted in the highest IE (least efficient performance). A similar pattern was also observed as the number of targets increased (Figures \ref{fig:IE}b and \ref{fig:IE}c). While \Voice resulted in lower IE for a smaller number of targets, \Pinch produced lower IE overall at a higher number of targets, outperforming both \Dwell and \VSelect (\autoref{fig:IE}c).

\begin{figure*}[t!]
    \centering
    \begin{subfigure}[t]{0.5\columnwidth}
        \includegraphics[height=1\linewidth]{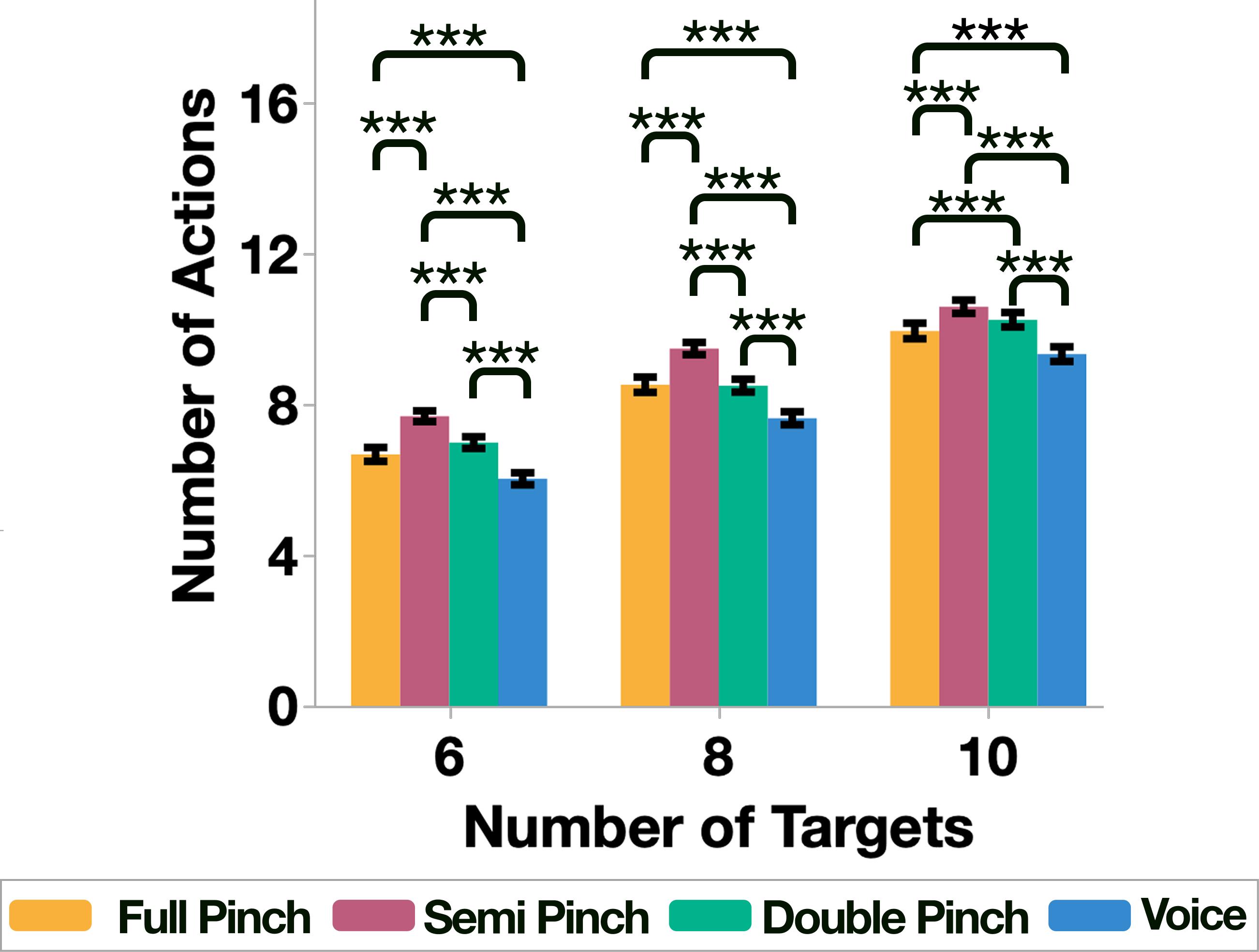}
        \caption{}
        \label{fig:action-target}
    \end{subfigure}
    \hspace{0.2\columnwidth}
    \begin{subfigure}[t]{0.5\columnwidth}
        \includegraphics[height=1\linewidth]{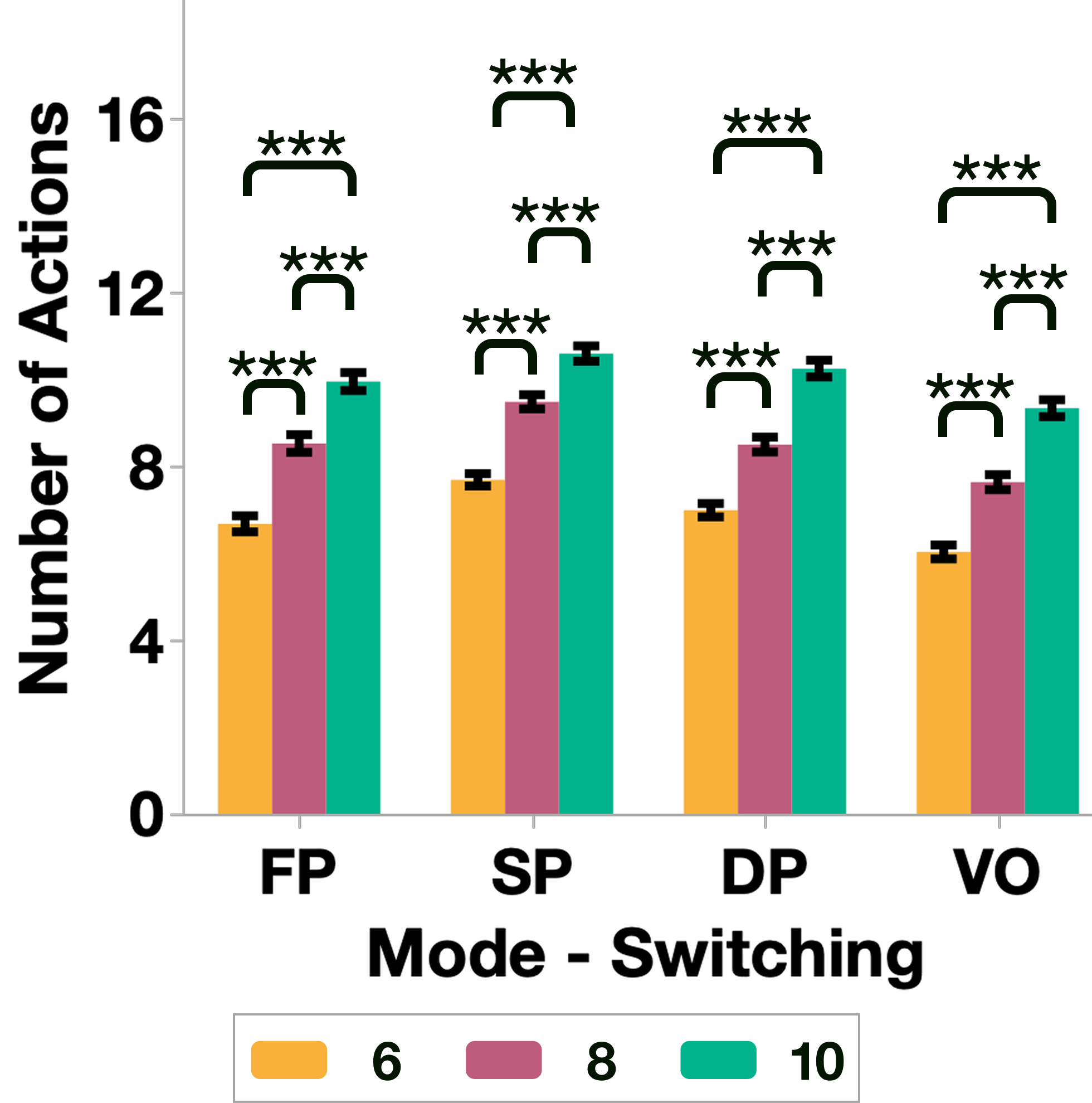}
        \caption{}
        \label{fig:action-mode-target}
    \end{subfigure}

    \caption{Number of actions (NoA) results for the interaction between (a) mode-switching and number of targets, and (b) subselection and mode-switching.} 
    \label{fig:action}
        \Description{Effect of mode-switching techniques, subselection techniques, and number of targets on mean number of actions. Number of actions performed across subselection techniques, switching techniques, and target set sizes. (a) Interaction of mode-switching technique and number of targets: required actions count scaled consistently with target number across all switching techniques, though DoublePinch and Voice required fewer actions overall compared to quasi-modes (FullPinch and SemiPinch).  (b) Number of actions by mode-switching techniques. The error bars represent 95\% confidence intervals. Brackets above bars indicate post-hoc significance: * p < .05, ** p < .01, and *** p < .001.}
\end{figure*}

\subsection{Number of Actions (NoA) -- \autoref{fig:action}}

Analysis of the NoA revealed a significant interaction between mode-switching and number of targets, $F(3.306,95.887) = 4.18, \, p < .006, \, \eta_p^2 = .097$. While all mode-switching techniques scaled with the target number, \Voice required the fewest actions, whereas \Semipinch consistently required the most actions. 

\begin{figure*}[!htb]
    \centering
    \includegraphics[width=1\linewidth]{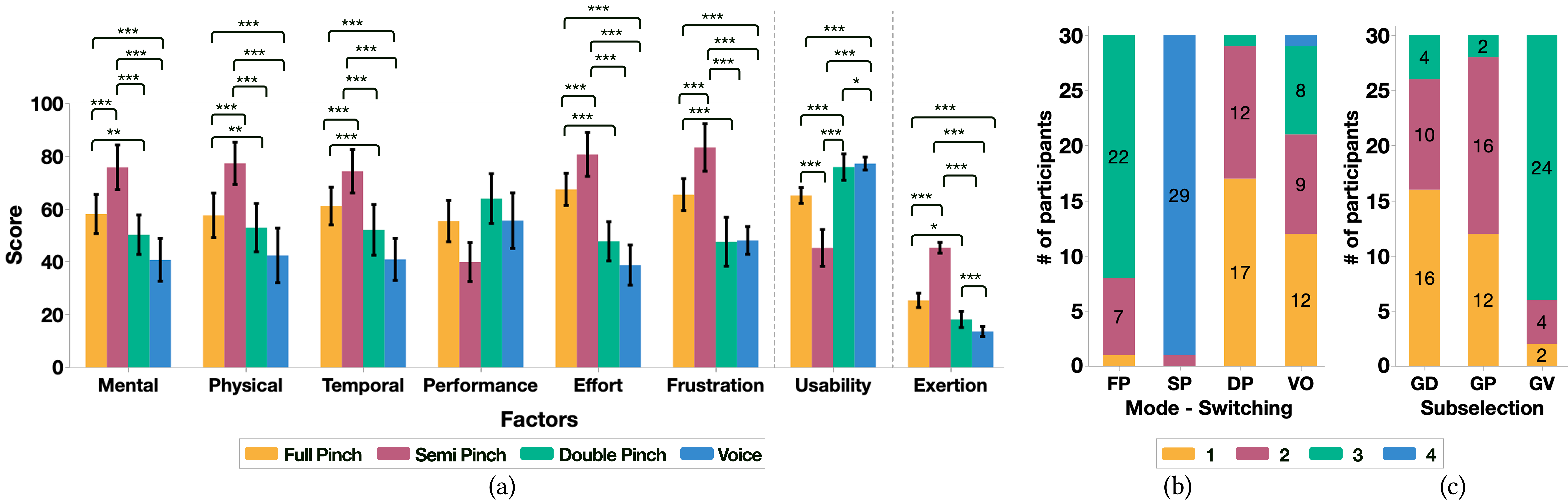}
    \caption{User experience results of the multi-selection techniques. Average (a) NASA-RTLX (Mental Demand, Physical Demand, Temporal Demand, Performance, Effort, Frustration), SUS (Usability), and the perceived Borg CR10 (Exertion) scores for the four mode-switching techniques; Borg CR10 scores were linearly scaled to 0–100 for consistency. Participants’ preference rankings for the (b) four mode-switching and (c) three subselection techniques.}   
    \label{fig:Subjective}
        \Description{The result of quantitative measures on the user experience of mode-switching and subselection techniques. (a) Mean NASA-TLX subscales (Mental, Physical, Temporal, Performance, Effort, Frustration), System Usability Scale (Usability), and perceived Exertion scores for the four mode-switching techniques; Borg CR10 exertion scores were linearly scaled to a 0–100 range for consistency. (b) Distribution of participants’ preference rankings (1 = best, 4 = worst) for the four mode-switching techniques. DoublePinch is most frequently ranked first, followed by Voice, whereas SemiPinch is most often ranked last. (c) Distribution of participants’ rankings for the three subselection techniques: Dwell (GD), Pinch (GP), Voice (GV). GV was ranked first place the most, while GD is most frequently ranked last. The error bars represent 95\% confidence intervals. Brackets above bars indicate post-hoc significance: * p < .05, ** p < .01, and *** p < .001.}
\end{figure*}

\subsection{Subjective Results -- \autoref{fig:Subjective}}\label{sec:subjective}

For Borg CR10, we found a significant effect of mode-switching, $\chi^2(3) = 66.24, p < .001$. Post-hoc analysis showed that \Semipinch caused the highest arm fatigue. \Fullpinch was also rated more fatiguing than \Doublepinch, while \Voice was consistently perceived as requiring the least effort (\autoref{fig:Subjective}a). Usability ratings (SUS) also differed significantly, $\chi^2(3) = 80.19, p < .001$. \Doublepinch achieved the best scores, followed by \Voice, \Fullpinch, with \Semipinch being the least usable. NASA-RTLX results revealed clear differences in workload, $\chi^2(3) = 75.05, p < .001$. Post-hoc analysis showed that \Semipinch demanded the most effort, followed by \Fullpinch, while \Doublepinch and \Voice were significantly less demanding. Frustration and temporal demand followed the same pattern. Physical demand was higher for \Semipinch and \Fullpinch than for \Doublepinch, while \Voice was consistently rated to demand the least workload. 

Preferences mirrored these trends. As shown in Figures \ref{fig:Subjective}b and \ref{fig:Subjective}c, 17 participants (56.7\%) ranked \Doublepinch as their top choice and 12 (40\%) favored \Voice, while \Semipinch was least preferred by most. For subselection, 16 (53.3\%) participants preferred \Dwell, 12 (40\%) chose \Pinch as the best, and 16 (53.3\%) ranked it the second-best, whereas only 2 (6.67\%) ranked \VSelect as most preferred, noticing \VSelect's hands-free nature but also describing it as tiring and inconsistent. Detailed participant feedback for each technique is presented below:

\textit{\Fullpinch.}
Participants found \Fullpinch familiar. Several noted it ``\textit{felt natural}'' (P6, P10, P21), but four reported that sustaining the pinch gesture became ``\textit{tiring}'' (P7, P18, P22, P29). P13 complained, ``\textit{holding my fingers tightly made them numb},'' while P20 mentioned it was ``\textit{fine for a short time but uncomfortable when longer.}''

\textit{\Semipinch.}
\Semipinch was consistently criticized as ``\textit{uncertain}'' (P1, P12, P16, P28), ``\textit{not reliable}'' (P4, P14, P22, P24), and ``\textit{hard to keep}'' (P4, P19, P23). Users reported frequent unintentional mode switches, such as P30, ``\textit{it was sometimes going back to single mode without my intention},'' and P17, ``\textit{I was watching my fingers instead of the objects}.'' Participants (P11, P22) also described it as the most fatiguing gesture. While they appreciated the mode status feedback (\autoref{fig:grid}), they had to focus on the gesture to sustain it. P11 said, ``\textit{I needed to pay extra attention to it cause it was not stable}'', with P19 summarizing it as ``\textit{the least usable}.'' Only P21 described it as ``\textit{interesting}'' and ``\textit{comfortable}''. 

\textit{\Doublepinch.}
Most users who ranked \Doublepinch first described it as ``\textit{clear}'' (P3, P9, P14, P15) and ``\textit{easy}'' (P8, P24, P26, P27), finding that it provided a strong sense of control. P27 emphasized that the two quick pinches were ``\textit{easy to perform without thinking}'' and the mode status feedback (\autoref{fig:grid}) was helpful, which contributed to greater confidence in use. Two participants (P11, P25) reported \Doublepinch being slow at first but became faster once they ``\textit{got the hang of it.}'' Two (P2, P5) related the gesture to a mouse double-click, finding the similarity ``\textit{interesting}.''

\textit{\Voice.}
Feedback on \Voice was largely positive. Several participants valued its low physical effort, calling it ``\textit{comfortable}'' (P2, P8, P24) and ``\textit{easy to use}'' (P15, P26). The flexibility to choose from multiple words was appreciated, with P5 commenting, ``\textit{I could just say the word that felt natural to me.}'' A couple of them (P12, P29) mentioned occasional recognition issues or felt ``\textit{awkward speaking aloud}.'' One participant (P24) remarked that it ``\textit{felt similar to interacting with my HomePod.}'' 

\textit{\Dwell.}
Feedback on \Dwell was mixed. Some participants appreciated its ``\textit{effortless}'' nature (P1, P10, P18), with P7 saying, ``\textit{I didn’t need to move my hands at all, it was nice.}'' Others, however criticized it as ``\textit{slow}'' (P5, P16) and prone to unintended activations. Multiple participants (P5, P14, P20, P22, P30) reported that ``\textit{sometimes it selected when I didn’t want it to},'' while P27 described experiencing eye fatigue from prolonged fixations. 

\textit{\Pinch.}
Participants generally described \Pinch as ``\textit{fast}'' (P4, P11, P17) and ``\textit{precise}'' (P6, P13, P19). However, few (P15, P28) remarked on finger fatigue, particularly when many targets had to be selected while maintaining a \Semipinch with the NDH. P23 reported that while holding \Fullpinch on the NDH, they occasionally released it unintentionally when performing a pinch with the DH. We made similar observations during the study for other participants as well.

\textit{\VSelect.}
Opinions on \VSelect were mixed as well. Some participants highlighted its hands-free convenience, calling it ``\textit{relaxing}'' (P3, P12) and ``\textit{less tiring}'' (P27). P8, however, expressed their concerns, ``\textit{It was nice not having to move my hands, but saying it over and over is not pleasant},'' while P26 noted that ``\textit{it worked best when I just said the same word each time.}'' Others (P15, P22, P29) criticized it as ``\textit{slow}'' or ``\textit{awkward},'' with P19 adding that ``\textit{sometimes it didn’t catch my command correctly.}'' P25 remarked ``\textit{Selecting all orange objects would have been better to say than selecting them one by one.}''

\section{Discussion}

In this work, we examined the combined effects of mode-switching and subselection techniques on multi-selection performance within gaze-based, controller-free XR UIs. We compared two \qmode mode-switching techniques (\Fullpinch, \Semipinch) with two \pmode alternatives (\Doublepinch, \Voice), together with three subselection techniques, Gaze+Dwell (\Dwell), Gaze+Pinch (\Pinch), and Gaze+ Voice (\VSelect), to multi-select 6, 8, or 10 targets (\autoref{fig:teaser}). Our findings reveal a trade-off between these strategies, particularly in terms of mode stability, error cost, and cognitive demand.

\subsection{RQ1: \qmode vs. \pmode Mode-Switching}

\pmode-modes consistently outperformed \qmode-modes across TCT, ME, ER, NoA, workload, usability, and user preference. Although \qmode-modes offer faster mode-switching (\autoref{fig:mode-switch}), their interaction costs were much higher. A key weakness of \qmode-modes was their instability. Small fluctuations in muscle tension, natural tremor, or brief reductions in hand-tracking confidence frequently caused unperceived mode exits. Particularly for \Semipinch, even momentary unintentional relaxation resulted in reverting the mode to single-selection. The next subselection thereby cleared the entire previously selected group, thus requiring regrouping of the whole set of targets, i.e., an error with a global performance cost. Thus, \textit{\Semipinch showed higher error rates, steeper increases in TCT, higher cognitive load and fatigue (\autoref{fig:Subjective}a), and the lowest subjective preference ratings (\autoref{fig:Subjective}b)}. 

In contrast, \Fullpinch was comparatively more stable as maintaining the gesture was more natural and easy due to the inherent support and tactile feedback of the two fingers~\cite{james2022multi}. However, it was still perceived as fatiguing due to the need to sustain the posture. Also, we intentionally did not design \Fullpinch to revert to single-selection mode upon tracking loss and instead ended the trial. We did this as reverting modes would have conflicted with our design of ending the trial on pinch release, thereby making the interaction unnatural. While our design choices for \Semipinch elevated errors, we argue that choosing to end the trial early due to gesture misrecognition issues would have had similar outcomes. Although ME and NoA would have reduced, ER would have skyrocketed due to an increase in missed targets and a decrease in total grouped objects. Overall, \textit{both \qmode techniques suffered from substantial cognitive and physical effort} as users needed to continuously monitor their finger posture while simultaneously trying to gaze at targets, disrupting flow, and increasing attentional demands. Simply, the \qmode gestures were a constant distraction. 

\pmode-modes, by contrast, were more immune to tracking issues as there was no need to sustain the gesture/command. Further, forgetting to change the mode at the beginning of a trial, a common drawback of \pmode-mode~\cite{sellen1992prevention}, was less costly than unintentional gesture-tracking loss, as the error was immediately visible and local: selecting the second object in the wrong (i.e., single-selection) mode reverted the selection of only the previously selected item, visually reminding users of their mistake. These reasons potentially explain why \textbf{\pmode-mode techniques, \Doublepinch and \Voice, offered a more robust and efficient performance}, with lower TCT and error rates, even as the number of targets increased, and why they were rated by participants as more usable, reliable, and less tiring. Moreover, these findings also align with Norman’s principles~\cite{norman1983design}: \pmode toggles make state legible, actions reversible, and feedback immediate, thereby narrowing the \textit{``gulfs of execution and evaluation''}.

\subsection{RQ2: Scaling Behavior of Pinch-Based Mode-Switching}

Prior work~\cite{PinchCatcher} used DH semi-pinch mode-switching with NDH pinch subselection for small target sets (2–6 items). In contrast, we assigned \Semipinch to the NDH (based on prior findings that NDH gestures tend to perform better~\cite{smith2019experimental, Smith2020EvaluatingTS, surale2019experimental}) and \Pinch subselection to the DH~\cite{guiard1987asymmetric}. Further, as per our design of \Semipinch, and as discussed above, the momentary gesture-tracking losses produced a snowball effect as target counts increased --- every unintended exit forced regrouping from scratch, raising frustration and fatigue, and thus, making further slips more likely. Although it is unclear how Kim et al.~\cite{PinchCatcher} dealt with gesture-tracking loss, we believe that these design differences of DH vs. NDH gesture assignment, different target set sizes, and accommodating a more realistic cost of mode slips in our study potentially explain why \Semipinch+ \Pinch yielded higher errors in our work (7.1\%; \autoref{fig:ER}c) compared to the $\approx$2\% reported previously~\cite{PinchCatcher}. 

Moreover, our participants completed the 6-target condition task faster (7333 ms TCT for \Semipinch+ \Pinch) compared to $\approx$9000 ms in previous work~\cite{PinchCatcher}. One possible explanation is that continuously selecting a larger number of targets (6, 8, or 10) leads to faster selection of each individual object on average compared to selecting fewer targets (e.g., 2 items), as more repetitive actions enable participants to adopt a more efficient rhythm. Yet, and in line with speed-accuracy trade-offs~\cite{Batmaz2022}, this faster interaction may have further contributed to the higher errors observed in our study.

Overall, our results suggest a potentially interesting phenomenon that extends prior work~\cite{PinchCatcher}: \textit{\Semipinch is viable for short, hand-focused interactions with a small set of targets, but is vulnerable when sustained over longer sequences or under noisy hand-tracking conditions}. Yet, further work is required to verify this.

As for \Fullpinch, it was more stable than \Semipinch but continued to decline with an increase in the number of targets. In contrast, \textbf{\Doublepinch achieved superior performance and scaled more efficiently} (\autoref{fig:IE}). 

\subsection{RQ3: Interplay Between Mode-Switching and Subselection}

Mode-switching stability directly influenced subselection performance. When \qmode-modes slipped, users had to split attention between maintaining the gesture and confirming targets, disrupting the natural flow of gaze-based selection. In contrast, \pmode-modes provided more efficient performance (\autoref{fig:IE}) as mode maintenance required no attention, and thus, subselection techniques operated more independently.

Across subselection techniques, users were fastest with \Pinch, but it incurred more errors, potentially due to late-triggers ~\cite{kumar2008improving, GazeHandSync} when used in quick successions. \Dwell was noted as more ``\textit{effortless}'' but ``\textit{slow}'' and error-prone \cite{Jacob1995Midas}. \VSelect reduced physical effort but added speech recognition latency, monotony, and occasional recognition errors. Participants disliked repeating short voice commands for each object, and performance was slower than \Pinch (\autoref{fig:TCT}). Instead, \VSelect was perceived to be better suited for high-level commands rather than granular, item-by-item confirmations.

Combinations further clarified these trade-offs. \textbf{\Doublepinch+ \Pinch produced the fastest, most accurate, and preferred interaction}, further supporting bimanual interaction and reinforcing the dominance of gaze+pinch interaction in recent XR devices~\cite{pfeuffer2024design, APV, GXR}. \Voice+ \VSelect was reliable but slow. Pairing \VSelect with \qmode-mode techniques slightly reduced subselection errors (\autoref{fig:ER}a), but the inherent instability of \qmode-modes still required frequent regrouping.

Although our study used a controlled layout, the patterns we observed --- \qmode-mode instability, \Doublepinch+ \Pinch outperforming other alternatives, and voice recognition latency --- come from intrinsic properties of the input techniques, not the spatial arrangement. We validated this in a separate pilot study with 12 participants using varied target sizes ($1.72^{\circ}$, $2.86^{\circ}$) and a better 3D representation of the 2D target grid (while keeping the rest of the experimental setup identical to the main study). There, we found performance patterns to be consistent with our main user study (see supplementary materials). We therefore expect these underlying trade-offs to potentially hold for similar 3D environments, though factors such as depth and occlusion need further investigation.

\section{Design Recommendations} 

\textbf{Favor \pmode over \qmode strategies for serial multi-selection of moderate target sets.}
\pmode-mode toggles (\Doublepinch, \Voice) sustained lower error rates (\autoref{fig:ER}) and better efficiency (\autoref{fig:IE}) for moderate target sets (6–10 targets). We expect these benefits to also extend to even larger sets as well, as \pmode strategies keep the state active until explicitly changed and avoid unintentional silent mode drops. Still, to improve mode awareness, it is critical that clear visual, auditory, or haptic cues make the active mode perceivable at all times.

\textbf{Use Gaze+Pinch (\Pinch) for granular, sequential selections.} The combination of gaze for targeting and pinch for confirmation delivered the best speed–accuracy trade-off \cite{mutasim2021pinch}. Pinch gestures remain natural and familiar for users.

\textbf{Limit reliance on dwell and repeated voice subselections.} While dwell reduces physical effort, it suffers from the Midas Touch problem \cite{Jacob1995Midas}. For voice, repeated commands were rated monotonous and socially awkward. Both techniques should be used with caution for sequential object selection and are more appropriate in scenarios involving fewer targets or where hands-free interaction is required, e.g., for people with limited muscle control \cite{mott2017improving}.

\textbf{Leverage voice for parallel or high-level actions.} Voice input is well-suited for ``one-to-many'' commands, such as confirming an entire highlighted group at once or for switching modes. Several participants noted that voice would be more compelling if it could act on multiple items with one/two utterances (e.g., ``select all,'' ``add these,'' or ``start'' $\rightarrow$ swipe with gaze $\rightarrow$ ``end'' \cite{hedeshy2021hummer}). Thus, we recommend designers avoid requiring a spoken command for each object and instead use voice to compress multiple actions into one/two steps.

\textbf{Consider target set size when selecting multi-selection techniques.} Although prior work~\cite{PinchCatcher} found semi-pinch a viable option for small target sets (2-6), our results showed that both \qmode techniques, \Fullpinch and \Semipinch, degraded sharply as the number of targets increased from 6 to 10. In contrast, \pmode-modes scaled more efficiently, and thus are better suited for scenarios involving larger groups of objects.

\section{Limitations and Future Work}
Here, we focused on serial multi-selection with up to ten targets in a controlled grid layout. Although this reflects common XR contexts such as file or photo management, real-world scenarios may involve denser, irregular, or 3D spatial distributions of objects. Thus, our findings should be interpreted as a controlled baseline rather than a complete account of all 3D contexts. Future work should extend our evaluation to more naturalistic 3D layouts, including varying depth planes and spatial occlusions, to assess how the trade-offs observed in our work transfer to more complex environments. Similarly, we recommend future studies to investigate parallel selection techniques (e.g., volume~\cite{wu2023point}) that may have benefits, particularly for densely arranged items. 

Moreover, our focus in this work was to isolate and evaluate the core multi-selection mechanism. Thus, we intentionally scoped out deselection to avoid introducing additional confounds. Although deselection is an important component of multi-selection workflows, understanding its interaction with serial and parallel multi-selection requires further investigation. Future work should therefore examine how different deselection strategies influence multi-selection performance in XR.

Finally, our experimental task involved targets and distractors differentiated by color to ensure experimental control. While this facilitated systematic comparisons, we acknowledge that this design choice diverges from real-world XR applications where targets are not pre-highlighted, and users must visually search for and classify objects based on attributes such as name, size, content of an image, or spatial location.

\section{Conclusion}

Here, we examined multi-selection in XR and how different mode-switching and subselection strategies influence user performance and experience. By comparing two \qmode techniques (\Fullpinch, \Semipinch) with two \pmode techniques (\Doublepinch, \Voice), and pairing them with three subselection methods (Gaze+Pinch, Gaze+Dwell, Gaze+Voice), to multi-select 6-10 targets, we provided \textit{the first} comprehensive evaluation of this design space.

Our results show that \pmode strategies, particularly \Doublepinch combined with Gaze+Pinch subselection, yielded the most efficient performance, while \qmode strategies suffered from instability, fatigue, and high error rates. Hands-free \Voice-based switching was well received and reduced effort, yet repeated voice commands for subselection were less preferred compared to hand-gesture confirmation. 
These findings highlight that mode-switching stability is critical for multi-selection. More broadly, the results emphasize the interplay between mode-switching and subselection --- when mode-switching is unstable, subselection suffers, amplifying errors and workload. Based on these findings, we provided design recommendations that could potentially influence multi-selection in XR and beyond.

\bibliographystyle{ACM-Reference-Format}
\bibliography{references}

@String{Computing = "Computing" }

@String{Computer = "{IEEE} Computer" }

@String{Springer = "Springer-Verlag" }

@book{laviola20173d,
  title={3D user interfaces: theory and practice},
  author={LaViola Jr, Joseph J and Kruijff, Ernst and McMahan, Ryan P and Bowman, Doug and Poupyrev, Ivan P},
  year={2017},
  publisher={Addison-Wesley Professional}
}

@inproceedings{wu2023point,
  title={Point-and volume-based multi-object acquisition in vr},
  author={Wu, Zhiqing and Yu, Difeng and Goncalves, Jorge},
  booktitle={IFIP Conference on Human-Computer Interaction},
  pages={20--42},
  year={2023},
  organization={Springer}
}

@phdthesis{lucas2005design,
  title={Design and evaluation of 3D multiple object selection techniques},
  author={Lucas, John Finley},
  year={2005},
  school={Virginia Tech}
}

@inproceedings{chen2020disambiguation,
  title={Disambiguation techniques for freehand object manipulations in virtual reality},
  author={Chen, Di Laura and Balakrishnan, Ravin and Grossman, Tovi},
  booktitle={2020 IEEE conference on virtual reality and 3D user interfaces (VR)},
  pages={285--292},
  year={2020},
  organization={IEEE}
}

@article{sellen1992prevention,
  title={The prevention of mode errors through sensory feedback},
  author={Sellen, Abigail J and Kurtenbach, Gordon P and Buxton, William AS},
  journal={Human-computer interaction},
  volume={7},
  number={2},
  pages={141--164},
  year={1992},
  publisher={Taylor \& Francis}
}

@phdthesis{ramanathan2016multi,
  title={Multi-Selection in Touch-Screen Graphical User Interfaces},
  author={Ramanathan, Anirudh},
  year={2016}
}

@inproceedings{pfeuffer2014gaze,
  title={Gaze-touch: combining gaze with multi-touch for interaction on the same surface},
  author={Pfeuffer, Ken and Alexander, Jason and Chong, Ming Ki and Gellersen, Hans},
  booktitle={Proceedings of the 27th annual ACM symposium on User interface software and technology},
  pages={509--518},
  year={2014}
}

@inproceedings{dehmeshki2010design,
  title={Design and evaluation of a perceptual-based object group selection technique},
  author={Dehmeshki, Hoda and Stuerzlinger, Wolfgang},
  booktitle={Proceedings of HCI 2010},
  year={2010},
  organization={BCS Learning \& Development}
}

@inproceedings{hinckley2006springboard,
  title={The springboard: multiple modes in one spring-loaded control},
  author={Hinckley, Ken and Guimbretiere, Francois and Baudisch, Patrick and Sarin, Raman and Agrawala, Maneesh and Cutrell, Ed},
  booktitle={Proceedings of the SIGCHI conference on Human Factors in computing systems},
  pages={181--190},
  year={2006}
}

@article{shi2024experimental,
  title={Experimental analysis of freehand multi-object selection techniques in virtual reality head-mounted displays},
  author={Shi, Rongkai and Wei, Yushi and Hu, Xuning and Liu, Yu and Yue, Yong and Yu, Lingyun and Liang, Hai-Ning},
  journal={Proceedings of the ACM on Human-Computer Interaction},
  volume={8},
  number={ISS},
  pages={93--111},
  year={2024},
  publisher={ACM New York, NY, USA}
}

@article{zhang2023multi,
  title={Multi-finger-based arbitrary region-of-interest selection in virtual reality},
  author={Zhang, Qimeng and Kim, Chang-Hun and Byun, Hae Won},
  journal={International Journal of Human--Computer Interaction},
  volume={39},
  number={20},
  pages={3969--3983},
  year={2023},
  publisher={Taylor \& Francis}
}

@inproceedings{pfeuffer2017gaze+,
author = {Pfeuffer, Ken and Mayer, Benedikt and Mardanbegi, Diako and Gellersen, Hans},
title = {Gaze + pinch interaction in virtual reality},
year = {2017},
isbn = {9781450354868},
publisher = {Association for Computing Machinery},
address = {New York, NY, USA},
url = {https://doi.org/10.1145/3131277.3132180},
doi = {10.1145/3131277.3132180},
booktitle = {Proceedings of the 5th Symposium on Spatial User Interaction},
pages = {99–108},
numpages = {10},
location = {Brighton, United Kingdom},
series = {SUI '17}
}

@inproceedings{PinchCatcher,
author = {Kim, Jinwook and Park, Sangmin and Zhou, Qiushi and Gonzalez-Franco, Mar and Lee, Jeongmi and Pfeuffer, Ken},
title = {PinchCatcher: Enabling Multi-selection for Gaze+Pinch},
year = {2025},
isbn = {9798400713941},
publisher = {Association for Computing Machinery},
address = {New York, NY, USA},
url = {https://doi.org/10.1145/3706598.3713530},
doi = {10.1145/3706598.3713530},
booktitle = {Proceedings of the 2025 CHI Conference on Human Factors in Computing Systems},
articleno = {853},
numpages = {16},
location = {
},
series = {CHI '25}
}

@inproceedings{smith2019experimental,
  title={Experimental Analysis of Single Mode Switching Techniques in Augmented Reality.},
  author={Smith, Jesse and Wang, Isaac and Woodward, Julia and Ruiz, Jaime},
  booktitle={Graphics Interface},
  pages={20--1},
  year={2019}
}

@inproceedings{surale2019experimental,
  title={Experimental analysis of barehand mid-air mode-switching techniques in virtual reality},
  author={Surale, Hemant Bhaskar and Matulic, Fabrice and Vogel, Daniel},
  booktitle={Proceedings of the 2019 CHI conference on human factors in computing systems},
  pages={1--14},
  year={2019}
}

@inproceedings{bergstrom2021evaluate,
  title={How to evaluate object selection and manipulation in vr? guidelines from 20 years of studies},
  author={Bergstr{\"o}m, Joanna and Dalsgaard, Tor-Salve and Alexander, Jason and Hornb{\ae}k, Kasper},
  booktitle={proceedings of the 2021 CHI conference on human factors in computing systems},
  pages={1--20},
  year={2021}
}

@misc{hair2014multivariate,
  title={Multivariate data analysis},
  author={Hair Jr, Joseph F and Black, William C and Babin, Barry J and Anderson, Rolph E.},
  year={2014},
  publisher={Pearson Education Limited}
  
}

@article{mallery2003spss,
  title={{SPSS} for Windows step by step: a simple guide and reference},
  author={Mallery, Paul and George, Darren},
  journal={Allyn, Bacon, Boston,},
  year={2003}
}

@inproceedings{hart2006nasa,
  title={NASA-task load index (NASA-TLX); 20 years later},
  author={Hart, Sandra G},
  booktitle={Proceedings of the human factors and ergonomics society annual meeting},
  volume={50},
  number={9},
  pages={904--908},
  year={2006},
  organization={Sage publications Sage CA: Los Angeles, CA}
}

@article{brooke2013sus,
  title={SUS: a retrospective.},
  author={Brooke, John},
  journal={Journal of usability studies},
  volume={8},
  number={2},
  year={2013}
}

@article{statsenko2020applying,
  title={Applying the inverse efficiency score to visual--motor task for studying speed-accuracy performance while aging},
  author={Statsenko, Yauhen and Habuza, Tetiana and Gorkom, Klaus Neidl-Van and Zaki, Nazar and Almansoori, Taleb M},
  journal={Frontiers in Aging Neuroscience},
  volume={12},
  pages={574401},
  year={2020},
  publisher={Frontiers Media SA}
}

@inproceedings{wentzel2020improving,
  title={Improving virtual reality ergonomics through reach-bounded non-linear input amplification},
  author={Wentzel, Johann and d'Eon, Greg and Vogel, Daniel},
  booktitle={Proceedings of the 2020 CHI Conference on Human Factors in Computing Systems},
  pages={1--12},
  year={2020}
}

@inproceedings{hu2023gaze,
  title={Gaze-based mode-switching to enhance interaction with menus on tablets},
  author={Hu Fleischhauer, Yanfei and Surale, Hemant Bhaskar and Alt, Florian and Pfeuffer, Ken},
  booktitle={Proceedings of the 2023 Symposium on Eye Tracking Research and Applications},
  pages={1--8},
  year={2023}
}

@article{rocchi2025comparison,
  title={A comparison of the Meta Quest Pro and HTC Vive Focus 3 eye-tracking systems: analysis of data accuracy and spatial precision},
  author={Rocchi, Erica and Ferrarotti, Anna and Carli, Marco},
  journal={IEEE Access},
  year={2025},
  publisher={IEEE}
}

@article{bashar2025effect,
  title={The effect of visual depth on the vergence--accommodation conflict on 3D selection performance within virtual reality headsets: MR Bashar et al.},
  author={Bashar, Mohammad Raihanul and Barrera Machuca, Mayra Donaji and Stuerzlinger, Wolfgang and Batmaz, Anil Ufuk},
  journal={The Visual Computer},
  pages={1--17},
  year={2025},
  publisher={Springer}
}

@article{wang2025exploring,
  title={Exploring Confirmation Strategies for Voice Interaction in Multi-Tasking Scenario},
  author={Wang, Junfeng and Xu, Zhiyu and Zhai, Weimin and Xu, Fei},
  journal={International Journal of Social Robotics},
  pages={1--14},
  year={2025},
  publisher={Springer}
}

@inproceedings{schmitt2021voice,
  title={Voice as a contemporary frontier of interaction design},
  author={Schmitt, Anuschka and Zierau, Naim and Janson, Andreas and Leimeister, Jan Marco},
  booktitle={European Conference on Information Systems (ECIS).-Virtual},
  year={2021}
}

@article{fernandez2022multi,
  title={A multi-object grasp technique for placement of objects in virtual reality},
  author={Fern{\'a}ndez, Unai J and Elizondo, Sonia and Iriarte, Naroa and Morales, Rafael and Ortiz, Amalia and Marichal, Sebastian and Ardaiz, Oscar and Marzo, Asier},
  journal={Applied Sciences},
  volume={12},
  number={9},
  pages={4193},
  year={2022},
  publisher={MDPI}
}

@inproceedings{liebling2014gaze,
  title={Gaze and mouse coordination in everyday work},
  author={Liebling, Daniel J and Dumais, Susan T},
  booktitle={Proceedings of the 2014 ACM international joint conference on pervasive and ubiquitous computing: adjunct publication},
  pages={1141--1150},
  year={2014}
}

@inproceedings{piumsomboon2017exploring,
  title={Exploring natural eye-gaze-based interaction for immersive virtual reality},
  author={Piumsomboon, Thammathip and Lee, Gun and Lindeman, Robert W and Billinghurst, Mark},
  booktitle={2017 IEEE symposium on 3D user interfaces (3DUI)},
  pages={36--39},
  year={2017},
  organization={IEEE}
}

@inproceedings{mutasim2021pinch,
  title={Pinch, click, or dwell: Comparing different selection techniques for eye-gaze-based pointing in virtual reality},
  author={Mutasim, Aunnoy K and Batmaz, Anil Ufuk and Stuerzlinger, Wolfgang},
  booktitle={Acm symposium on eye tracking research and applications},
  pages={1--7},
  year={2021}
}

@inproceedings{jacob1990you,
  title={What you look at is what you get: eye movement-based interaction techniques},
  author={Jacob, Robert JK},
  booktitle={Proceedings of the SIGCHI conference on Human factors in computing systems},
  pages={11--18},
  year={1990}
}

@inproceedings{penkar2012designing,
  title={Designing for the eye: design parameters for dwell in gaze interaction},
  author={Penkar, Abdul Moiz and Lutteroth, Christof and Weber, Gerald},
  booktitle={Proceedings of the 24th australian computer-human interaction conference},
  pages={479--488},
  year={2012}
}

@inproceedings{isomoto2022interaction,
  title={Interaction design of dwell selection toward gaze-based ar/vr interaction},
  author={Isomoto, Toshiya and Yamanaka, Shota and Shizuki, Buntarou},
  booktitle={2022 Symposium on Eye Tracking Research and Applications},
  pages={1--2},
  year={2022}
}

@Inbook{majaranta2014eye,
    author="Majaranta, P{\"a}ivi and Bulling, Andreas",
    commented-editor="Fairclough, Stephen H. and Gilleade, Kiel",
    title="Eye Tracking and Eye-Based Human--Computer Interaction",
    bookTitle="Advances in Physiological Computing",
    year="2014",
    publisher="Springer London",
    address="London",
    pages="39--65",
    isbn="978-1-4471-6392-3",
    doi="10.1007/978-1-4471-6392-3_3",
    url="https://doi.org/10.1007/978-1-4471-6392-3_3"
}

@inproceedings{jang2017modeling,
  title={Modeling cumulative arm fatigue in mid-air interaction based on perceived exertion and kinetics of arm motion},
  author={Jang, Sujin and Stuerzlinger, Wolfgang and Ambike, Satyajit and Ramani, Karthik},
  booktitle={Proceedings of the 2017 CHI conference on human factors in computing systems},
  pages={3328--3339},
  year={2017}
}

@article{pfeuffer2024design,
  title={Design principles and challenges for gaze+ pinch interaction in xr},
  author={Pfeuffer, Ken and Gellersen, Hans and Gonzalez-Franco, Mar},
  journal={IEEE Computer Graphics and Applications},
  volume={44},
  number={03},
  pages={74--81},
  year={2024},
  publisher={IEEE Computer Society}
}

@inproceedings{lystbaek2024hands,
  title={Hands-on, Hands-off: Gaze-Assisted Bimanual 3D Interaction},
  author={Lystb{\ae}k, Mathias N and Mikkelsen, Thorbj{\o}rn and Krisztandl, Roland and Gonzalez, Eric J and Gonzalez-Franco, Mar and Gellersen, Hans and Pfeuffer, Ken},
  booktitle={Proceedings of the 37th Annual ACM Symposium on User Interface Software and Technology},
  pages={1--12},
  year={2024}
}

@inproceedings{chatterjee2015gaze+,
  title={Gaze+ gesture: Expressive, precise and targeted free-space interactions},
  author={Chatterjee, Ishan and Xiao, Robert and Harrison, Chris},
  booktitle={Proceedings of the 2015 ACM on international conference on multimodal interaction},
  pages={131--138},
  year={2015}
}

@inproceedings{pfeuffer2020empirical,
  title={Empirical evaluation of gaze-enhanced menus in virtual reality},
  author={Pfeuffer, Ken and Mecke, Lukas and Delgado Rodriguez, Sarah and Hassib, Mariam and Maier, Hannah and Alt, Florian},
  booktitle={Proceedings of the 26th ACM Symposium on Virtual Reality Software and Technology},
  pages={1--11},
  year={2020}
}

@inproceedings{shi2021exploring,
  title={Exploring head-based mode-switching in virtual reality},
  author={Shi, Rongkai and Zhu, Nan and Liang, Hai-Ning and Zhao, Shengdong},
  booktitle={2021 IEEE International Symposium on Mixed and Augmented Reality (ISMAR)},
  pages={118--127},
  year={2021},
  organization={IEEE}
}

@article{park2020analysis,
  title={Analysis of control characteristics between dominant and non-dominant hands by transient responses of circular tracking movements in 3D virtual reality space},
  author={Park, Wookhyun and Choi, Woong and Jo, Hanjin and Lee, Geonhui and Kim, Jaehyo},
  journal={Sensors},
  volume={20},
  number={12},
  pages={3477},
  year={2020},
  publisher={MDPI}
}

@article{bahill1975main,
  title={The main sequence, a tool for studying human eye movements},
  author={Bahill, A Terry and Clark, Michael R and Stark, Lawrence},
  journal={Mathematical biosciences},
  volume={24},
  number={3-4},
  pages={191--204},
  year={1975},
  publisher={Elsevier}
}

@inproceedings{hansen2018fitts,
  title={A Fitts' law study of click and dwell interaction by gaze, head and mouse with a head-mounted display},
  author={Hansen, John Paulin and Rajanna, Vijay and MacKenzie, I Scott and B{\ae}kgaard, Per},
  booktitle={Proceedings of the Workshop on Communication by Gaze Interaction},
  pages={1--5},
  year={2018}
}

@inproceedings{majaranta2009fast,
  title={Fast gaze typing with an adjustable dwell time},
  author={Majaranta, P{\"a}ivi and Ahola, Ulla-Kaija and {\v{S}}pakov, Oleg},
  booktitle={Proceedings of the sigchi conference on human factors in computing systems},
  pages={357--360},
  year={2009}
}

@inproceedings{mott2017improving,
  title={Improving dwell-based gaze typing with dynamic, cascading dwell times},
  author={Mott, Martez E and Williams, Shane and Wobbrock, Jacob O and Morris, Meredith Ringel},
  booktitle={Proceedings of the 2017 chi conference on human factors in computing systems},
  pages={2558--2570},
  year={2017}
}

@article{esteves2020comparing,
  title={Comparing selection mechanisms for gaze input techniques in head-mounted displays},
  author={Esteves, Augusto and Shin, Yonghwan and Oakley, Ian},
  journal={International Journal of Human-Computer Studies},
  volume={139},
  pages={102414},
  year={2020},
  publisher={Elsevier}
}

@article{pai2019assessing,
  title={Assessing hands-free interactions for VR using eye gaze and electromyography},
  author={Pai, Yun Suen and Dingler, Tilman and Kunze, Kai},
  journal={Virtual Reality},
  volume={23},
  number={2},
  pages={119--131},
  year={2019},
  publisher={Springer}
}

@inproceedings{zhao2022eyesaycorrect,
  title={Eyesaycorrect: Eye gaze and voice based hands-free text correction for mobile devices},
  author={Zhao, Maozheng and Huang, Henry and Li, Zhi and Liu, Rui and Cui, Wenzhe and Toshniwal, Kajal and Goel, Ananya and Wang, Andrew and Zhao, Xia and Rashidian, Sina and others},
  booktitle={Proceedings of the 27th International Conference on Intelligent User Interfaces},
  pages={470--482},
  year={2022}
}

@article{parisay2021eyetap,
  title={EyeTAP: Introducing a multimodal gaze-based technique using voice inputs with a comparative analysis of selection techniques},
  author={Parisay, Mohsen and Poullis, Charalambos and Kersten-Oertel, Marta},
  journal={International Journal of Human-Computer Studies},
  volume={154},
  pages={102676},
  year={2021},
  publisher={Elsevier}
}

@article{dondi2023gaze,
  title={Gaze-based human--computer interaction for museums and exhibitions: technologies, applications and future perspectives},
  author={Dondi, Piercarlo and Porta, Marco},
  journal={Electronics},
  volume={12},
  number={14},
  pages={3064},
  year={2023},
  publisher={MDPI}
}

@article{khan2021gavin,
  title={GAVIN: Gaze-assisted voice-based implicit note-taking},
  author={Khan, Anam Ahmad and Newn, Joshua and Kelly, Ryan M and Srivastava, Namrata and Bailey, James and Velloso, Eduardo},
  journal={ACM Transactions on Computer-Human Interaction (TOCHI)},
  volume={28},
  number={4},
  pages={1--32},
  year={2021},
  publisher={ACM New York, NY, USA}
}

@inproceedings{mutasim2025there,
  title={There Is More to Dwell Than Meets the Eye: Toward Better Gaze-Based Text Entry Systems With Multi-Threshold Dwell},
  author={Mutasim, Aunnoy K and Bashar, Mohammad Raihanul and Lutteroth, Christof and Batmaz, Anil Ufuk and Stuerzlinger, Wolfgang},
  booktitle={Proceedings of the 2025 CHI Conference on Human Factors in Computing Systems},
  pages={1--18},
  year={2025}
}

@inproceedings{10.1145/1842993.1842997,
author = {Wigdor, Daniel},
title = {Architecting next-generation user interfaces},
year = {2010},
isbn = {9781450300766},
publisher = {Association for Computing Machinery},
address = {New York, NY, USA},
url = {https://doi-org.lib-ezproxy.concordia.ca/10.1145/1842993.1842997},
doi = {10.1145/1842993.1842997},
booktitle = {Proceedings of the International Conference on Advanced Visual Interfaces},
pages = {16–22},
numpages = {7},
location = {Roma, Italy},
series = {AVI '10}
}

@article{nishida2023single,
  title={Single-tap latency reduction with single-or double-tap prediction},
  author={Nishida, Naoto and Ikematsu, Kaori and Sato, Junichi and Yamanaka, Shota and Tsubouchi, Kota},
  journal={Proceedings of the ACM on Human-Computer Interaction},
  volume={7},
  number={MHCI},
  pages={1--26},
  year={2023},
  publisher={ACM New York, NY, USA}
}

@article{owen2025improving,
  title={Improving Speech Recognition Accuracy in Noisy Environments Using Vosk and Custom Acoustic Models},
  author={Owen, Anthony and Orsten, Paul},
  year={2025}
}

@inproceedings{wills1996selection,
  title={Selection: 524,288 ways to say" this is interesting"},
  author={Wills, Graham J},
  booktitle={Proceedings IEEE Symposium on Information Visualization'96},
  pages={54--60},
  year={1996},
  organization={IEEE}
}

@article{Smith2020EvaluatingTS,
  title={Evaluating the Scalability of Non-Preferred Hand Mode Switching in Augmented Reality},
  author={Jesse Smith and Isaac Wang and Winston Wei and Julia Woodward and Jaime Ruiz},
  journal={Proceedings of the 2020 International Conference on Advanced Visual Interfaces},
  year={2020},
  url={https://api.semanticscholar.org/CorpusID:222110409}
}

@article{Pfeuffer2015GazeShiftingDI,
  title={Gaze-Shifting: Direct-Indirect Input with Pen and Touch Modulated by Gaze},
  author={Ken Pfeuffer and Jason Alexander and Ming Ki Chong and Yanxia Zhang and Hans-Werner Gellersen},
  journal={Proceedings of the 28th Annual ACM Symposium on User Interface Software \& Technology},
  year={2015},
  url={https://api.semanticscholar.org/CorpusID:15305806}
}

@article{Zhao2017RealtimeHG,
  title={Real-time head gesture recognition on head-mounted displays using cascaded hidden Markov models},
  author={Jingbo Zhao and Robert S. Allison},
  journal={2017 IEEE International Conference on Systems, Man, and Cybernetics (SMC)},
  year={2017},
  pages={2361-2366},
  url={https://api.semanticscholar.org/CorpusID:24959302}
}

@article{Oh2021A3H,
  title={A 3D head pointer: a manipulation method that enables the spatial
 position and posture for supernumerary robotic limbs},
  author={Joi Oh and Fumihiro Kato and Iwasaki Yukiko and Hiroyasu Iwata},
  journal={ACTA IMEKO},
  year={2021},
  url={https://api.semanticscholar.org/CorpusID:240085378}
}

@article{Jayasri2023ABR,
  title={A Brief Review on Voice Assisted Hands-free Virtual Painting System},
  author={A.R. Jayasri and Dr.P. Kanimozhi and Dr. T. Ananth Kumar},
  journal={Middle East Journal of Applied Science \& Technology},
  year={2023},
  url={https://api.semanticscholar.org/CorpusID:265347068}
}

@inproceedings{yi2022gazedock,
  title={Gazedock: Gaze-only menu selection in virtual reality using auto-triggering peripheral menu},
  author={Yi, Xin and Lu, Yiqin and Cai, Ziyin and Wu, Zihan and Wang, Yuntao and Shi, Yuanchun},
  booktitle={2022 IEEE Conference on Virtual Reality and 3D User Interfaces (VR)},
  pages={832--842},
  year={2022},
  organization={IEEE}
}

@online{google_tiltbrush_tools_online,
  author        = {{Google}},
  title         = {Using the Tilt Brush Tools, Quick Tools, and Menu panels},
  organization  = {Tilt Brush Help},
  url           = {https://support.google.com/tiltbrush/answer/6389713?hl=en},
  urldate       = {2025-09-05}
}

@online{shapelab_site_online,
  title         = {Shapelab — Official Website},
  organization  = {Leopoly},
  url           = {https://shapelabvr.com/},
  urldate       = {2025-09-05}
}

@online{mennuti_blocks_next_level_2018,
  author        = {Brittany Mennuti},
  title         = {Take your Blocks models to the next level},
  date          = {2018-01-29},
  url           = {https://blog.google/products/google-ar-vr/take-your-blocks-models-next-level/},
  organization  = {Google},
  journaltitle  = {The Keyword},
  urldate       = {2025-09-05}
}

@misc{appleSelectItems,
  author       = {Apple Inc.},
  title        = {Select items on your Mac screen},
  howpublished = {\url{https://support.apple.com/en-ca/guide/mac-help/mchlp1378/mac}},
  note         = {Mac User Guide — macOS Sequoia 15},
  urldate      = {2025-09-06},
  year         = {2025}
}

@misc{APV,
  author       = {Apple Inc.},
  title        = {Learn basic gestures and controls on Apple Vision Pro},
  howpublished = {\url{https://support.apple.com/en-ca/guide/apple-vision-pro/tan1e2a29e00/visionos}},
  urldate      = {2025-12-01},
  year         = {2025}
}

@misc{GXR,
  author       = {Galaxy XR},
  title        = {Galaxy XR Interaction},
  howpublished = {\url{https://www.samsung.com/us/xr/galaxy-xr/galaxy-xr/}},
  urldate      = {2025-12-01},
  year         = {2025}
}

@inproceedings{schon2023tailor,
  title={Tailor twist: Assessing rotational mid-air interactions for augmented reality},
  author={Sch{\"o}n, Dominik and Kosch, Thomas and M{\"u}ller, Florian and Schmitz, Martin and G{\"u}nther, Sebastian and Bommhardt, Lukas and M{\"u}hlh{\"a}user, Max},
  booktitle={Proceedings of the 2023 CHI Conference on Human Factors in Computing Systems},
  pages={1--14},
  year={2023}
}

@inproceedings{hou2023classifying,
  title={Classifying head movements to separate head-gaze and head gestures as distinct modes of input},
  author={Hou, Baosheng James and Newn, Joshua and Sidenmark, Ludwig and Ahmad Khan, Anam and B{\ae}kgaard, Per and Gellersen, Hans},
  booktitle={Proceedings of the 2023 CHI Conference on Human Factors in Computing Systems},
  pages={1--14},
  year={2023}
}

@article{souchet2023narrative,
  title={A narrative review of immersive virtual reality’s ergonomics and risks at the workplace: cybersickness, visual fatigue, muscular fatigue, acute stress, and mental overload},
  author={Souchet, Alexis D and Lourdeaux, Domitile and Pagani, Alain and Rebenitsch, Lisa},
  journal={Virtual Reality},
  volume={27},
  number={1},
  pages={19--50},
  year={2023},
  publisher={Springer}
}

@article{palmisano2020cybersickness,
  title={Cybersickness in head-mounted displays is caused by differences in the user's virtual and physical head pose},
  author={Palmisano, Stephen and Allison, Robert S and Kim, Juno},
  journal={Frontiers in Virtual Reality},
  volume={1},
  pages={587698},
  year={2020},
  publisher={Frontiers Media SA}
}

@inproceedings{bragdon2011gesture,
  title={Gesture select: acquiring remote targets on large displays without pointing},
  author={Bragdon, Andrew and Ko, Hsu-Sheng},
  booktitle={Proceedings of the SIGCHI Conference on Human Factors in Computing Systems},
  pages={187--196},
  year={2011}
}

@inproceedings{hyrskykari2012gaze,
  title={Gaze gestures or dwell-based interaction?},
  author={Hyrskykari, Aulikki and Istance, Howell and Vickers, Stephen},
  booktitle={Proceedings of the Symposium on Eye Tracking Research and Applications},
  pages={229--232},
  year={2012}
}

@inproceedings{zhai1999manual,
  title={Manual and gaze input cascaded (MAGIC) pointing},
  author={Zhai, Shumin and Morimoto, Carlos and Ihde, Steven},
  booktitle={Proceedings of the SIGCHI conference on Human factors in computing systems},
  pages={246--253},
  year={1999}
}

@inproceedings{pfeuffer2023palmgazer,
  title={Palmgazer: Unimanual eye-hand menus in augmented reality},
  author={Pfeuffer, Ken and Obernolte, Jan and Dietz, Felix and M{\"a}kel{\"a}, Ville and Sidenmark, Ludwig and Manakhov, Pavel and Pakanen, Minna and Alt, Florian},
  booktitle={Proceedings of the 2023 ACM Symposium on Spatial User Interaction},
  pages={1--12},
  year={2023}
}

@inproceedings{bashar2025depth3dsketch,
  title={Depth3DSketch: Freehand Sketching Out of Arm's Reach in Virtual Reality},
  author={Bashar, Mohammad Raihanul and Amini, Mohammadreza and Stuerzlinger, Wolfgang and Sarac, Mine and Pfeuffer, Ken and Barrera Machuca, Mayra Donaji and Batmaz, Anil Ufuk},
  booktitle={Proceedings of the Extended Abstracts of the CHI Conference on Human Factors in Computing Systems},
  pages={1--8},
  year={2025}
}

@inproceedings{sidenmark2023vergence,
  title={Vergence matching: Inferring attention to objects in 3d environments for gaze-assisted selection},
  author={Sidenmark, Ludwig and Clarke, Christopher and Newn, Joshua and Lystb{\ae}k, Mathias N and Pfeuffer, Ken and Gellersen, Hans},
  booktitle={Proceedings of the 2023 CHI Conference on Human Factors in Computing Systems},
  pages={1--15},
  year={2023}
}

@inproceedings{casiez20121,
  title={1€ filter: a simple speed-based low-pass filter for noisy input in interactive systems},
  author={Casiez, G{\'e}ry and Roussel, Nicolas and Vogel, Daniel},
  booktitle={Proceedings of the SIGCHI Conference on Human Factors in Computing Systems},
  pages={2527--2530},
  year={2012}
}

@article{guiard1987asymmetric,
  title={Asymmetric division of labor in human skilled bimanual action: The kinematic chain as a model},
  author={Guiard, Yves},
  journal={Journal of motor behavior},
  volume={19},
  number={4},
  pages={486--517},
  year={1987},
  publisher={Taylor \& Francis}
}

@inproceedings{fashimpaur2023investigating,
  title={Investigating wrist deflection scrolling techniques for extended reality},
  author={Fashimpaur, Jacqui and Karlson, Amy and Jonker, Tanya R and Benko, Hrvoje and Gupta, Aakar},
  booktitle={Proceedings of the 2023 CHI Conference on Human Factors in Computing Systems},
  pages={1--16},
  year={2023}
}

@inproceedings{kumar2008improving,
  title={Improving the accuracy of gaze input for interaction},
  author={Kumar, Manu and Klingner, Jeff and Puranik, Rohan and Winograd, Terry and Paepcke, Andreas},
  booktitle={Proceedings of the 2008 symposium on Eye tracking research \& applications},
  pages={65--68},
  year={2008}
}

@inproceedings{mutasim2022saccades,
    author = {Mutasim, Aunnoy and Batmaz, Anil Ufuk and Hudhud Mughrabi, Moaaz and Stuerzlinger, Wolfgang},
    title = {Performance Analysis of Saccades for Primary and Confirmatory Target Selection},
    year = {2022},
    isbn = {9781450398893},
    publisher = {Association for Computing Machinery},
    address = {New York, NY, USA},
    url = {https://doi.org/10.1145/3562939.3565619},
    doi = {10.1145/3562939.3565619},
    booktitle = {Proceedings of the 28th ACM Symposium on Virtual Reality Software and Technology},
    articleno = {18},
    numpages = {12},
    location = {Tsukuba, Japan},
    series = {VRST '22}
}

@article{hirzle2022understanding,
  title={Understanding, addressing, and analysing digital eye strain in virtual reality head-mounted displays},
  author={Hirzle, Teresa and Fischbach, Fabian and Karlbauer, Julian and Jansen, Pascal and Gugenheimer, Jan and Rukzio, Enrico and Bulling, Andreas},
  journal={ACM Transactions on Computer-Human Interaction (TOCHI)},
  volume={29},
  number={4},
  pages={1--80},
  year={2022},
  publisher={ACM New York, NY}
}

@InProceedings{lu2020blinks,
    author={X. {Lu} and D. {Yu} and H. -N. {Liang} and W. {Xu} and Y. {Chen} and X. {Li} and K. {Hasan}},
    booktitle={2020 IEEE International Symposium on Mixed and Augmented Reality (ISMAR)},   title={Exploration of Hands-free Text Entry Techniques For Virtual Reality}, 
    publisher = {IEEE},
    year={2020},  
    volume={},  
    number={},  
    pages={344-349},  
    doi={10.1109/ISMAR50242.2020.00061}
}

@inproceedings{Sidenmark2019_2,
    author = {Sidenmark, Ludwig and Gellersen, Hans},
    title = {Eye\&Head: Synergetic Eye and Head Movement for Gaze Pointing and Selection},
    year = {2019},
    isbn = {9781450368162},
    publisher = {Association for Computing Machinery},
    address = {New York, NY, USA},
    url = {https://doi.org/10.1145/3332165.3347921},
    doi = {10.1145/3332165.3347921},
    booktitle = {Proceedings of the 32nd Annual ACM Symposium on User Interface Software and Technology},
    pages = {1161–1174},
    numpages = {14},
    location = {New Orleans, LA, USA},
    series = {UIST '19}
}

@misc{Patidar2014,
    title={Quickpie: An Interface for Fast and Accurate Eye Gazed based Text Entry}, 
    author={Pawan Patidar and Himanshu Raghuvanshi and Sayan Sarcar},
    year={2014},
    eprint={1407.7313},
    archivePrefix={arXiv},
    primaryClass={cs.HC}
}

@inproceedings{hedeshy2021hummer,
    author = {Hedeshy, Ramin and Kumar, Chandan and Menges, Raphael and Staab, Steffen},
    title = {Hummer: Text Entry by Gaze and Hum},
    year = {2021},
    isbn = {9781450380966},
    publisher = {Association for Computing Machinery},
    address = {New York, NY, USA},
    url = {https://doi.org/10.1145/3411764.3445501},
    doi = {10.1145/3411764.3445501},
    booktitle = {Proceedings of the 2021 CHI Conference on Human Factors in Computing Systems},
    articleno = {741},
    numpages = {11},
    location = {Yokohama, Japan},
    series = {CHI '21}
}

@article{norman1983design,
  title={Design rules based on analyses of human error},
  author={Norman, Donald A},
  journal={Communications of the ACM},
  volume={26},
  number={4},
  pages={254--258},
  year={1983},
  publisher={ACM New York, NY, USA}
}

@inproceedings{chaconas2018evaluation,
  title={An evaluation of bimanual gestures on the microsoft hololens},
  author={Chaconas, Nikolas and H{\"o}llerer, Tobias},
  booktitle={2018 IEEE conference on virtual reality and 3D user interfaces (VR)},
  pages={1--8},
  year={2018},
  organization={IEEE}
}

@inproceedings{hinrichs2011gestures,
  title={Gestures in the wild: studying multi-touch gesture sequences on interactive tabletop exhibits},
  author={Hinrichs, Uta and Carpendale, Sheelagh},
  booktitle={Proceedings of the SIGCHI Conference on Human Factors in Computing Systems},
  pages={3023--3032},
  year={2011}
}

@inproceedings{Rajanna2018,
    author = {Rajanna, Vijay and Hansen, John Paulin},
    title = {Gaze Typing in Virtual Reality: Impact of Keyboard Design, Selection Method, and Motion},
    year = {2018},
    isbn = {9781450357067},
    publisher = {Association for Computing Machinery},
    address = {New York, NY, USA},
    url = {https://doi.org/10.1145/3204493.3204541},
    doi = {10.1145/3204493.3204541},
    booktitle = {Proceedings of the 2018 ACM Symposium on Eye Tracking Research and Applications},
    articleno = {15},
    numpages = {10},
    location = {Warsaw, Poland},
    series = {ETRA '18}
}

@article{lang2023multimodal,
  title={A multimodal smartwatch-based interaction concept for immersive environments},
  author={Lang, Mat{\v{e}}j and Strobel, Clemens and Weckesser, Felix and Langlois, Danielle and Kasneci, Enkelejda and Kozl{\'\i}kov{\'a}, Barbora and Krone, Michael},
  journal={Computers \& Graphics},
  volume={117},
  pages={85--95},
  year={2023},
  publisher={Elsevier}
}

@inproceedings{yeo2019wrist,
  title={Wrist: Watch-ring interaction and sensing technique for wrist gestures and macro-micro pointing},
  author={Yeo, Hui-Shyong and Lee, Juyoung and Kim, Hyung-il and Gupta, Aakar and Bianchi, Andrea and Vogel, Daniel and Koike, Hideki and Woo, Woontack and Quigley, Aaron},
  booktitle={Proceedings of the 21st international conference on human-computer interaction with mobile devices and services},
  pages={1--15},
  year={2019}
}

@inproceedings{javaheri2025llms,
  title={LLMs Enable Context-Aware Augmented Reality in Surgical Navigation},
  author={Javaheri, Hamraz and Ghamarnejad, Omid and Lukowicz, Paul and Stavrou, Gregor Alexander and Karolus, Jakob},
  booktitle={Proceedings of the 2025 ACM Designing Interactive Systems Conference},
  pages={3205--3220},
  year={2025}
}

@article{valladares2025device,
  title={On-Device Automatic Speech Recognition for Low-Resource Languages in Mixed Reality Industrial Metaverse Applications: Practical Guidelines and Evaluation of a Shipbuilding Application in Galician},
  author={Valladares-Poncela, Ant{\'o}n and Fraga-Lamas, Paula and Fern{\'a}ndez-Caram{\'e}s, Tiago M},
  journal={IEEE Access},
  year={2025},
  publisher={IEEE}
}

@article{Mutasim2025GEM_Gaze,
    author = {Mutasim, Aunnoy K and Batmaz, Anil Ufuk and Hudhud Mughrabi, Moaaz and Stuerzlinger, Wolfgang},
    title = {The Influence of Eye Gaze Interaction Technique Expertise and the Guided Evaluation Method on Text Entry Performance Evaluations},
    year = {2025},
    issue_date = {May 2025},
    publisher = {Association for Computing Machinery},
    address = {New York, NY, USA},
    volume = {9},
    number = {3},
    url = {https://doi.org/10.1145/3725842},
    doi = {10.1145/3725842},
    journal = {Proc. ACM Hum.-Comput. Interact.},
    month = may,
    articleno = {ETRA17},
    numpages = {19}
}

@inproceedings{xie2024intelligent,
  title={Intelligent Voice Assistant Interaction Design in Virtual Reality Environments},
  author={Xie, Jing},
  booktitle={2024 IEEE 6th International Conference on Civil Aviation Safety and Information Technology (ICCASIT)},
  pages={60--64},
  year={2024},
  organization={IEEE}
}

@article{godden2025robotic,
  title={Robotic Characterization of Markerless Hand-Tracking on Meta Quest Pro and Quest 3 Virtual Reality Headsets},
  author={Godden, Eric and Steedman, William and Pan, Matthew KXJ},
  journal={IEEE Transactions on Visualization and Computer Graphics},
  year={2025},
  publisher={IEEE}
}

@article{monteiro2025beyond,
  title={Beyond the Hands: Evaluating the Usability of Hands-Free Methods and Controllers for Menu Selection During an Immersive VR Experience},
  author={Monteiro, Pedro and Peixoto, Bruno and Gon{\c{c}}alves, Guilherme and Coelho, Hugo and Barbosa, Lu{\'\i}s and Melo, Miguel and Bessa, Maximino},
  journal={International Journal of Human--Computer Interaction},
  pages={1--38},
  year={2025},
  publisher={Taylor \& Francis}
}

@inproceedings{GazeHandSync,
author = {Park, Yeji and Kim, Jiwan and Oakley, Ian},
title = {GazeHandSync: Mitigating Late-Trigger Errors for Seamless Gaze-Hand Interactions},
year = {2025},
isbn = {9798400714870},
publisher = {Association for Computing Machinery},
address = {New York, NY, USA},
url = {https://doi-org.lib-ezproxy.concordia.ca/10.1145/3715669.3723126},
doi = {10.1145/3715669.3723126},
booktitle = {Proceedings of the 2025 Symposium on Eye Tracking Research and Applications},
articleno = {12},
numpages = {8},
location = {
},
series = {ETRA '25}
}

@article{Choe2019,
    author = {Mungyeong Choe and Yeongcheol Choi and Jaehyun Park and Hyun K. Kim},
    title = {Comparison of Gaze Cursor Input Methods for Virtual Reality Devices},
    journal = {International Journal of Human–Computer Interaction},
    volume = {35},
    number = {7},
    pages = {620-629},
    year  = {2019},
    publisher = {Taylor \& Francis},
    doi = {10.1080/10447318.2018.1484054},
    
    URL = { 
            https://doi.org/10.1080/10447318.2018.1484054
        
    },
    eprint = { 
            https://doi.org/10.1080/10447318.2018.1484054
        
    }
}

@article{Jacob1995Midas,
    title={Eye tracking in advanced interface design},
    author={Jacob, Robert JK},
    journal={Virtual environments and advanced interface design},
    volume={258},
    number={288},
    pages={2},
    year={1995},
    publisher={Citeseer}
}

@ARTICLE{Batmaz2022,
  author={Batmaz, Anil Ufuk and Stuerzlinger, Wolfgang},
  journal={IEEE Transactions on Visualization and Computer Graphics}, 
  title={Effective Throughput Analysis of Different Task Execution Strategies for Mid-Air Fitts' Tasks in Virtual Reality}, 
  year={2022},
  volume={28},
  number={11},
  pages={3939-3947},
  doi={10.1109/TVCG.2022.3203105}
}

@article{james2022multi,
  title={Multi-Finger Haptics: Analysis of Human Hand Grasp towards a Tripod Three-Finger Haptic Grasp model},
  author={James, Jose},
  journal={arXiv preprint arXiv:2301.00049},
  year={2022}
}
\end{document}